%% file: fulleos.tex
\documentstyle[prd,aps,epsf]{revtex}

\def\simgt{\,\rlap{\lower 3.5 pt\hbox{$\mathchar \sim$}}\raise 1pt \hbox {$>$}\,}
\def\simlt{\,\rlap{\lower 3.5 pt\hbox{$\mathchar \sim$}}\raise 1pt \hbox {$<$}\,}
\begin{document}

\draft

\tightenlines

\title{
{\normalsize \hfill {\sf UTHEP-440}} \\
\vspace*{-2pt}
{\normalsize \hfill {\sf UTCCP-P-101}} \\
\vspace*{-2pt}
{\normalsize \hfill {\sf March, 2001}} \\
Equation of state in finite-temperature QCD with two flavors 
of improved Wilson quarks
}

\author{CP-PACS Collaboration :
  A.~Ali~Khan\rlap,$^{\rm a}$\thanks{address till 31 August, 2000}
  S.~Aoki\rlap,$^{\rm b}$
  R.~Burkhalter\rlap,$^{\rm a,b}$
  S.~Ejiri\rlap,$^{\rm a}$
  M.~Fukugita\rlap,$^{\rm c}$
  S.~Hashimoto\rlap,$^{\rm d}$
  N.~Ishizuka\rlap,$^{\rm a,b}$
  Y.~Iwasaki\rlap,$^{\rm a,b}$
  K.~Kanaya\rlap,$^{\rm b}$
  T.~Kaneko\rlap,$^{\rm d}$
  Y.~Kuramashi\rlap,$^{\rm d}$
  T.~Manke\rlap,$^{\rm a}$\thanks{present address:
        Department of Physics, Columbia University,
        538 W 120th St., New York, NY 10027, USA}
  K.-I.~Nagai\rlap,$^{\rm a}$
  M.~Okamoto\rlap,$^{\rm b}$
  M.~Okawa\rlap,$^{\rm d}$
  H.P.~Shanahan\rlap,$^{\rm a}$\thanks{present address:
        Department of Biochemistry and Molecular
        Biology, University College London, London, England, UK}
  Y.~Taniguchi\rlap,$^{\rm b}$
  A.~Ukawa$^{\rm a,b}$ and
  T.~Yoshi\'e$^{\rm a,b}$
 }

\address{
$^{\rm a}$Center for Computational Physics,
    University of Tsukuba, Tsukuba, Ibaraki 305-8577, Japan, \\
$^{\rm b}$Institute of Physics,
    University of Tsukuba, Tsukuba, Ibaraki 305-8571, Japan, \\
$^{\rm c}$Institute for Cosmic Ray Research,
    University of Tokyo, Kashiwa 277-8582, Japan, \\
$^{\rm d}$High Energy Accelerator Research Organization
    (KEK), Tsukuba, Ibaraki 305-0801, Japan}

\date{\today}
\maketitle

\begin{abstract}
We present results of a first study of equation of state 
in finite-temperature QCD with two flavors of Wilson-type quarks. 
Simulations are made on lattices with temporal size $N_t=4$ and 6, 
using an RG-improved action for the gluon sector and a meanfield-improved 
clover action for the quark sector. 
The lines of constant physics corresponding to fixed values of the ratio 
$m_{\rm PS}/m_{\rm V}$ of the pseudo-scalar to vector meson masses at 
zero temperature are determined, and 
the beta functions which describe the renormalization-group flow along 
these lines are calculated. 
Using these results, 
the energy density and the pressure are calculated as functions of 
temperature along the lines of constant physics in the range 
$m_{\rm PS}/m_{\rm V} = 0.65$--0.95. 
The quark mass dependence in the equation of state is found to be small 
for $m_{\rm PS}/m_{\rm V} \simlt 0.8$. 
Comparison of results for $N_t=4$ and $N_t=6$ lattices show significant 
scaling violation present in the $N_t=4$ results. 
At high temperatures the results for $N_t=6$ are quite close to 
the continuum Stefan-Boltzmann limit, suggesting the possibility of 
a precise continuum extrapolation of thermodynamic quantities from 
simulations at $N_t\simgt 6$. 
\end{abstract}



\newpage

\section{Introduction}
\label{sec:intro}

During the last decade, much effort has been devoted to experimentally 
detecting the quark-gluon plasma state in high energy heavy-ion 
collisions. 
In order to extract an unambiguous signal of quark-gluon plasma production 
from heavy-ion collision experiments, theoretical understanding 
on the nature of the finite-temperature chiral phase transition and 
the thermodynamic properties of quark-gluon plasma is indispensable. 
In particular, the equation of state (EOS) belongs to the most basic 
category of information needed in phenomenological investigations of 
heavy-ion collisions. 

Extensive numerical studies have been pursued in lattice QCD to derive 
the equation of state from first principles \cite{ejiri00}. 
Within the quenched approximation in which effects of dynamical quark-pair 
creation and annihilation are neglected, 
precise results have been established. 
Continuum extrapolations of the lattice results have been made with various 
lattice actions, finding a good agreement within errors of a few 
percent\cite{Boy96,okamoto,Bei99}. 
For the pressure, a detailed comparison of the results from 
the integral method and the derivative method  
have also been made \cite{Kla98,Eji98,Eng99}.
The problem of non-zero pressure gap at the transition point with 
the derivative method has been solved by a non-perturbative 
calculation of anisotropy coefficients \cite{Eji98}. 

The next step towards a realistic quark-gluon plasma simulation is 
to include dynamical 
quarks, which clearly play a significant role in the real world 
through chiral symmetry. 
Until recently, EOS with dynamical quarks has been computed only with 
the Kogut-Susskind (staggered) quark action or its improved form 
\cite{Ber97b,Eng97,Kar00}.
Strictly speaking, the staggered quark action only allows 
the number of flavors to be a multiple of four, and providing 
a mass difference within the four-fold multiplet is not straightforward. 
It has also been found that the critical scaling for two-flavor QCD 
extracted with this formalism \cite{KL94,JLQCD98,Laermann98,MILC99} 
does not agree with the theoretically expected $O(4)$ values. 
These features of the Kogut-Susskind quark action make it imperative 
that the quark-gluon plasma properties be explored with alternative quark 
actions.  In this article we present results on the equation of state 
obtained with the Wilson quark action in an improved form. 

Study of finite temperature QCD with Wilson-type quark action has been 
difficult for two reasons.  First, explicit chiral symmetry breaking 
complicates the phase diagram analysis \cite{Fuku88,Ber92,Iwa96}, 
which is basic for obtaining the equation of state. 
In this connection, 
an important role played by the parity-flavor broken phase \cite{aoki} 
has been realized, and the phase structure for finite temporal 
lattice sizes has been understood \cite{AUU,AKUU,AIKKUY}. 

Another difficulty has been that, when the standard plaquette gauge action 
and the standard Wilson quark action are adopted, 
the system exhibits severe lattice artifacts on coarse lattices 
with the temporal lattice size $N_t=4$ and 6.  For example, 
the finite-temperature transition strengthens at intermediate quark masses 
to a first-order transition for $N_t=6$ \cite{Ber92,Iwa96}, 
while it should weaken as quarks become heavier.
In this regard, it has been shown that improvement of the gauge action 
is effective in reducing the lattice artifacts in finite temperature QCD. 
Furthermore, the critical scaling around the two-flavor chiral 
transition obtained for a renormalization-group (RG) improved 
gauge action is consistent with the expected $O(4)$ universality 
class at $N_t=4$ \cite{Iwa97PRL}. 

These advances indicate that thermodynamic studies with Wilson-type 
quark actions are feasible if improved actions are employed.  
We have thus attempted a first calculation of EOS in QCD with two flavors of 
dynamical quarks, using an RG improved gauge action \cite{iwasaki} 
coupled with a clover-improved Wilson quark action \cite{SWclover}.
This combination of actions is motivated from our previous 
comparative study \cite{CPPACScom}, 
where we found lattice discretization errors to be small 
with this action combination in both gluonic and hadronic observables 
at zero temperature. 

The phase structure and the critical temperature 
for this action combination have been studied in Ref.~\cite{nf2FT} 
employing an $N_t=4$ lattice. 
In this paper, we extend the study to an $N_t=6$ lattice.  We then 
calculate the pressure and energy density for $N_t=4$ and 6 
employing the integral method \cite{Eng90}. 
We obtain EOS as a function of temperature 
for each fixed value of the renormalized quark mass, 
{\it i.e.,} on each line of constant physics. 
We identify these lines by the ratio of pseudo-scalar and vector meson masses, 
$m_{\rm PS}/m_{\rm V}$, at zero temperature.
Our results covers EOS over the range $m_{\rm PS}/m_{\rm V} = 0.65$--0.95. 

The organization of the paper is as follows. 
Our lattice action and the simulation parameters are summarized 
in Sec.~\ref{sec:model}. 
In Sec.~\ref{sec:phase}, we discuss
the phase structure of QCD for our improved Wilson quark action 
at $N_t=4$ and 6.
In Sec.~\ref{sec:scale}, the temperature scale and the lines of 
constant physics are studied. 
The RG beta functions, required in our calculation of EOS, are 
determined in Sec.~\ref{sec:betafnc}. 
Results for EOS at $N_t=4$ and 6 are presented in Sec.~\ref{sec:eos}. 
Conclusions and discussions are given in Sec.~\ref{sec:concl}.

\section{Action and simulation parameters}
\label{sec:model}

The gluon \cite{iwasaki} and quark \cite{SWclover} action we employ 
is defined by 
\begin{eqnarray}
 S & = & S_g + S_q, 
\label{eq:action} \\
 S_g & = & -\beta \{ \sum_{x,\, \mu > \nu} 
      c_{0} W_{\mu \nu}^{1 \times 1}(x) 
    + \sum_{x,\, \mu, \nu} c_{1} W_{\mu \nu}^{1 \times 2}(x) \},
\label{eq:gaction} \\
S_q & = & \sum_{f=1,2} \sum_{x,y}\overline{q}_x^f D_{x,y} q_y^f.
\label{eq:qaction}
\end{eqnarray}
Here $\beta=6/g^2$, $c_1=-0.331$, $c_0=1-8c_1$ and 
\begin{eqnarray}
D_{x,y} & = & \delta_{xy}
- K \sum_\mu \{(1-\gamma_\mu)U_{x,\mu} \delta_{x+\hat\mu,y}
      + (1+\gamma_\mu)U_{x,\mu}^{\dag} \delta_{x,y+\hat\mu} \}
- \, \delta_{xy} c_{SW} K \sum_{\mu > \nu}
         \sigma_{\mu\nu} F_{\mu\nu}.
\end{eqnarray}
with $F_{\mu\nu}$ the lattice field strength,
\begin{equation}
F_{\mu\nu} = \frac{1}{8i} (f_{\mu\nu} - f_{\mu\nu}^{\dag}),
\label{eq:CloverLeaf}
\end{equation}
where $f_{\mu\nu}$ is the standard clover-shaped combination of gauge links. 
For the clover coefficient $c_{SW}$, 
we adopt a mean-field value $c_{SW} = P^{-3/4}$
substituting the one-loop result for the plaquette 
$P=1-0.8412 \beta^{-1}$ \cite {iwasaki}, 
which agrees within 8\% with the values measured in our runs \cite{CPPACSmq}. 
We also note that the one-loop result $c_{SW}=1+0.678/\beta+\cdots$ 
\cite{aoki99} is close to our choice $c_{SW}=1+0.631/\beta+\cdots$.

Our two-flavor simulation employs the standard HMC algorithm. 
Details of the algorithm implementation are the same as 
in Refs.~\cite{CPPACScom,nf2FT,CPPACSmq}.   
The length of one trajectory is unity, and the molecular dynamics time step 
is chosen to yield an acceptance rate greater than about 80\%. 
The inversion of quark matrix is made with the BiCGStab method.
We measure the Polyakov loop and its susceptibility at every trajectory.
Jack-knife errors of these expectation values are estimated with a bin size 
of 20--50 trajectories.
Hadron propagators are measured at every 5 trajectories using point and 
exponentially smeared quark sources.

In Ref.~\cite{nf2FT}, we studied the phase structure for our action combination 
on $16^3 \times 4$ lattices with a temporal lattice size $N_t=4$.
The simulation parameters are summarized in Table~\ref{tab:param16x4}. 
The values of the hopping parameter $K$ cover the range 
$m_{\rm PS}/m_{\rm V} \approx 0.60$--0.98.
In these simulations, we have also measured the observables required for 
EOS. 
We have since extended the simulation to an $N_t=6$ lattice. 
Simulation parameters for $N_t=6$ are compiled in Table~\ref{tab:param16x6}. 
For the spatial lattice size, we choose $N_s=16$ both for $N_t=4$ and 6. 
This enables us to commonly apply results obtained on a $16^4$ 
lattice to carry out the zero-temperature subtraction in the calculations 
of EOS, and to determine the lines of constant physics. 
For various tests, we also perform simulations on $8^3 \times 4$ lattices
as summarized in Appendix~\ref{sec:app1}. 

Parameters for the zero temperature simulations are compiled 
in Table~\ref{tab:param16x16}.
On the zero temperature lattice $16^4$, 
we determine meson masses by a combined fit using both point and smeared 
sources assuming a double hyperbolic cosine form for the propagator.  
This procedure is necessitated by the fact that a plateau of effective mass 
is sometimes not quite clear due to a small temporal size of 16.  
Results for masses are summarized in Table~\ref{tab:mass} 
and plotted in Figs.~\ref{fig:mpi2} and \ref{fig:rho}. 

In order to check the accuracy of mass results, we compare the values 
of $m_{\rm PS}$ and $m_{\rm V}$ with those obtained in our previous 
high statistic simulations \cite{CPPACSmq}. 
We find that, when $N_t a$ is smaller than about $6/m_{\rm PS}$
($K=0.1365$ at $\beta=2.2$ and $K=0.1355$ and 0.1360 at $\beta=2.3$), 
several masses are inconsistent with those obtained on 
a $24^3 \times 48$ lattice.
In these cases, the effective mass on the $16^4$ lattice show no clear 
plateau up to the largest time separation 8, 
suggesting that the temporal lattice size of 16 is not sufficiently large 
to remove contamination from excited states. 
Therefore, we do not use data of $m_{\rm PS}$ and $m_{\rm V}$ 
on the $16^4$ lattice when $N_t a < 6.0 / m_{\rm PS}$. 
We instead include results from our previous study \cite{CPPACSmq}, 
shown by open symbols in Figs.~\ref{fig:mpi2} and \ref{fig:rho}, 
in the analyses in the present study.

Since the aspect ratio $N_s/N_t=16/6=2.666\cdots$ for $N_t=6$ is smaller than 
$N_s/N_t=4$ for $N_t=4$, we also check the influence of the spatial 
volume on EOS. 
In the ideal gas limit of $\beta\to\infty$, analytic calculations 
as described in Appendix \ref{sec:app2} show a 10\% finite size 
correction for $N_s/N_t=3$ when $N_t\sim 4$--6 as compared to a 5\% 
effect for $N_s/N_t=4$.  Perturbative estimates are not reliable 
close to the critical temperature, however.  To study the spatial 
volume effect in this region, we make additional simulations at 
$N_s/N_t=2$ on $8^3\times4$ 
lattices and compare the results with those at $N_s/N_t=4$ 
obtained on $16^3\times4$ lattices.
Details are presented in Appendix~\ref{sec:app1}. 
We find that the pressure at $N_s/N_t=2$ and 4 are consistent with each 
other within 1--2\% except very near the critical temperature. 
We therefore conclude that finite volume corrections are reasonably 
controlled for our $N_t=6$ lattices over the range of temperature we study. 

\section{Phase structure and pseudo-critical temperature}
\label{sec:phase}

Figure~\ref{fig:phase} summarize our results for the phase diagram. 
The solid line threading through filled circles denoted $K_c(T=0)$ is 
the location of the critical line 
where pion mass vanishes at zero temperature. 
It is the line of constant physics for massless quarks. 
Above the $K_c(T=0)$ line, parity-flavor symmetry 
of the Wilson-type quark action is broken spontaneously \cite{aoki,AKUU}. 
Pion mass vanishes along the line since the pion becomes the massless mode 
of a second-order transition associated with this spontaneous breakdown.
At zero temperature, the boundary of the parity-flavor broken phase 
(the $K_c(T=0)$ line) is expected to form a sharp cusp touching the free 
massless fermion point $K=1/8$ at $\beta=\infty$.

For finite temporal sizes $N_t$ corresponding to finite temperatures 
$T=(aN_t)^{-1}$, the parity-flavor broken phase retracts from the 
large $\beta$ limit, forming a cusp \cite{AUU,AKUU}.
The boundary of the parity-flavor broken phase at $N_t=4$, 
the $K_c(N_t=4)$ line, is shown by a thin line threading through open circles 
in Fig.~\ref{fig:phase}. 

The dashed line $K_t(N_t=4)$ through open diamonds represents 
the finite-temperature 
pseudo-critical line for a temporal size $N_t=4$, which is   
determined from the Polyakov loop and its susceptibility \cite{nf2FT}. 
The region to the right of $K_t$ (larger $\beta$) is the high 
temperature quark-gluon plasma phase, 
and that to the left of $K_t$ (smaller $\beta$) is 
the low temperature hadron phase. 
The crossing point of the $K_c(T=0)$ and $K_t(N_t=4)$ lines 
is the finite-temperature chiral phase transition point \cite{Iwa96}. 
As one observes in Fig.~\ref{fig:phase}, the chiral transition point is 
located close to the cusp of the parity-flavor broken phase, 
with the difference expected to be $O(a)$.
This is consistent with the picture that the massless pion, interpreted as 
the Nambu-Goldstone boson associated with spontaneous chiral symmetry breaking 
in the continuum limit, appears only in the cold phase. 

In Figs.~\ref{fig:pline6} and \ref{fig:suspl6} we present 
the expectation value of the Polyakov loop 
and its susceptibility obtained on a $16^3 \times 6$ lattice. 
We find a clear peak of the Polyakov loop susceptibility. 
Fitting the peak by a gaussian form using 3 or 4 points around the peak,
we determine $K_t(N_t=6)$ at $\beta = 2.0$, 2.1, 2.2 and 2.3, 
as summarized in Table~\ref{tab:Kt}. 
The results are shown by filled diamonds denoted as $K_t(N_t=6)$ in 
Fig.~\ref{fig:phase}.  

The pseudo-critical temperature $T_{pc}$ in units of the zero-temperature 
vector meson mass was studied in Ref.~\cite{nf2FT} for $N_t=4$. 
We repeat the study using the $N_t=6$ data. 
For this purpose, we interpolate the zero-temperature meson mass data 
to the $K_t(N_t=6)$ line by 
\begin{eqnarray}
(m_{\rm PS}a)^2 &=& B_{\rm PS}\left(\frac{1}{K}-\frac{1}{K_c}\right)
        +C_{\rm PS}\left(\frac{1}{K}-\frac{1}{K_c}\right)^2 \label{eq:mpi}, \\
m_{\rm V}a      &=& B_{\rm V}\left(\frac{1}{K}-\frac{1}{K_c}\right)
        +C_{\rm V}\left(\frac{1}{K}-\frac{1}{K_c}\right)^2 . \label{eq:mrho}
\end{eqnarray}
The values of $m_{\rm PS}/m_{\rm V}$ and $T_{pc}/m_{\rm V}$ at $K_t$ 
are summarized in Table~\ref{tab:Kt}.

Results of $T_{pc}/m_{\rm V}$ as a function of $(m_{\rm PS}/m_{\rm V})^2$ 
are plotted in Fig.~\ref{fig:tcmrho} for $N_t=4$ \cite{nf2FT} and 6.
We find that, in the range $m_{\rm PS}/m_{\rm V} = 0.65$--0.95 we study, 
values of $T_{pc}/m_{\rm V}$ at $N_t=4$ and 6 agree within about 10\%.

\section{Lines of constant physics}
\label{sec:scale}

In previous studies of EOS with staggered-type quark actions, 
the pressure and energy density are determined as functions of
temperature for a fixed value of bare quark mass $m_q^{(0)}a$. 
While $m_q^{(0)} a$ and $N_t$ are practically easy to set in simulations, 
the system at different temperatures (or values of $\beta$) will have 
different physical quark masses. 
This is not useful for phenomenological applications,  and we 
need to evaluate the temperature dependence of thermodynamic observables 
for a fixed renormalized quark mass, {\it i.e.,} on a line of constant physics. 

We identify the lines of constant physics in the parameter space 
$(\beta,K)$ by the values of $m_{\rm PS}/m_{\rm V}$ at zero temperature.
Deferring details of the interpolation procedure of hadron mass data 
to Sec.~\ref{sec:betafnc}, we show the lines of constant physics
corresponding to the values $m_{\rm PS}/m_{\rm V}=0.65$, 0.70, 0.75, 0.80,
0.85, 0.90, 0.95 and 0.975 by solid lines in Fig.~\ref{fig:Tcons}. 
In this figure, the bold line with open circles represents the critical 
line $K_c(T=0)$, corresponding to $m_{\rm PS}/m_{\rm V}=0$.
The bold lines with open diamonds and triangles are the $K_t$ lines
for $N_t=4$ and 6.

We also attempt to determine the lines of constant temperature. 
Here, we adopt the pseudo-critical temperature $T_{pc}$ 
on the same line of constant physics as the unit of temperature, 
where the temperature itself is determined through the zero-temperature 
vector meson mass $m_{\rm V}a$ as 
\begin{eqnarray}
\frac{T}{m_{\rm V}}(\beta, K) 
= \frac{1}{N_t \times m_{\rm V} a (\beta, K)}.
\end{eqnarray}
The ratio $T_{pc}/m_{\rm V}$ for the pseudo-critical temperature $T_{pc}$ is 
obtained by tuning $\beta$ and $K$ along the $K_t$ line
for each $N_t$.
We then interpolate $T_{pc}/m_{\rm V}$ as a function of $(m_{\rm PS}/m_{\rm V})^2$ 
by a Pade-type ansatz, 
\begin{eqnarray}
\label{eq:tpsfit}
\frac{T_{pc}}{m_{\rm V}} = A \ 
\frac{1 + B \, (m_{\rm PS}/m_{\rm V})^2}{1 + C \, (m_{\rm PS}/m_{\rm V})^2}.
\end{eqnarray} 
We obtain $A=0.2253(71), B=-0.933(17), C=-0.820(39)$ with 
$\chi^2/N_{df}=1.61/5$ for $N_t=4$, and 
$A=0.261(19)$, $B=-0.873(35)$, $C=-0.624(108)$ with 
$\chi^2/N_{df}=0.74/1$ for $N_t=6$. 
Dashed and dot-dashed lines shown in Fig.~\ref{fig:tcmrho} are 
the fit results for $N_t=4$ and 6.

The ansatz (\ref{eq:tpsfit}) does not incorporate the O(4) scaling 
behavior $T_{pc}(m_{\rm PS})=T_{pc}(0)+c\, m_{\rm PS}^{2/\beta\delta}$ 
with $1/\beta\delta=0.54$ expected close to the chiral limit or the constraint 
$T_{pc}=0$ in the heavy quark limit $m_{\rm PS}/m_{\rm V}=1$. 
Fits satisfying these constraints may be attempted, {\it e.g.}, by replacing 
$(m_{\rm PS}/m_{\rm V})^2$ with $(m_{\rm PS}/m_{\rm V})^{2/\beta\delta}$ 
and giving an additional factor $(1-(m_{\rm PS}/m_{\rm V})^2)$.  They yield  
curves which are close to that of (\ref{eq:tpsfit}) but agree less well 
with data.  Since we employ such fits for the purpose of interpolating 
the data for $T_{pc}/m_{\rm V}$ over the quark mass range 
$m_{\rm PS}/m_{\rm V} \simeq 0.65$--0.95 of our study, we choose to adopt 
(\ref{eq:tpsfit}) in the following analyses. 

For $N_t=6$, since the data for $T_{pc}/m_{\rm V}$ covers only 
the range $m_{\rm PS}/m_{\rm V} = 0.725$--0.972, we have to 
extrapolate the fit result down to $m_{\rm PS}/m_{\rm V} = 0.65$. 
We check the effect of extrapolation by performing a fit of $N_t=4$ 
data using only the points in the range 
$m_{\rm PS}/m_{\rm V} = 0.725$--0.972.
We find that the difference between this fit and our full fit 
is less than 1\% for $m_{\rm PS}/m_{\rm V} = 0.65$--0.7.

Finally, we normalize the temperature $T/m_{\rm V}$ by the pseudo-critical 
temperature $T_{pc}/m_{\rm V}$ on the same line of constant physics. 
Results of the procedure above for the lines of constant temperature are 
shown by dashed lines 
in Fig.~\ref{fig:Tcons} for $T/T_{pc} = 1.0$, 1.2, 1.4, 1.6, 1.8 and 2.0, 
where $K_t(N_t=4)$ is used to set $T_{pc}$. 
We observe that the $K_t$ line for $N_t=6$ is slightly discrepant from 
the $T/T_{pc} = 1.5$ line; this deviation represents scaling violation in 
$T_{pc}/m_{\rm V}$.

\section{Beta functions}
\label{sec:betafnc}

The renormalization group flow along the lines of constant physics 
is described by the beta functions. 
Their precise values are required in a calculation of 
the energy density $\epsilon / T^4$ discussed in Sec.~\ref{sec:eos}. 
In this section, we study the beta functions
\begin{eqnarray}
\left. a \frac{\partial \beta}{\partial a} 
\right|_{\frac{m_{\rm PS}}{m_{\rm V}}},
\hspace{10mm} 
\left. a \frac{\partial K}{\partial a} 
\right|_{\frac{m_{\rm PS}}{m_{\rm V}}}
\end{eqnarray}
for fixed values of $m_{\rm PS}/m_{\rm V}$, 
using results for $m_{\rm PS} a$ and $m_{\rm V} a$ at zero temperature. 

Since $m_{\rm V}$ is a physical quantity which we can take as independent 
of the lattice spacing $a$, the derivatives 
$a \frac{\partial \beta}{\partial a}$ and 
$a \frac{\partial K}{\partial a}$ with fixed $m_{\rm PS}/m_{\rm V}$ 
can be replaced by $m_{\rm V} a \frac{\partial \beta}{\partial (m_{\rm V} a)}$ and 
$m_{\rm V} a \frac{\partial K}{\partial (m_{\rm V} a)}$.
Naively, these quantities may be determined in the following way. 
First, one fits the values of $m_{\rm PS} a$ and $m_{\rm V} a$ measured 
at each $(\beta, K)$ by a suitable fit function, 
and differentiate the function in terms of $\beta$ and $K$.
The derivatives $\frac{\partial \beta}{\partial (m_{\rm V} a)}$ and 
$\frac{\partial K}{\partial (m_{\rm V} a)}$ can be calculated by solving 
\begin{eqnarray}
\left( \begin{array}{cc}
\frac{\partial \beta}{\partial (m_{\rm V}a)} &
\frac{\partial K}{\partial (m_{\rm V}a)} \\ 
\frac{\partial \beta}{\partial (m_{\rm PS}/m_{\rm V})} &
\frac{\partial K}{\partial (m_{\rm PS}/m_{\rm V})} 
\end{array} \right) = 
\left( \begin{array}{cc}
 \frac{\partial (m_{\rm V}a)}{\partial \beta} &
 \frac{\partial (m_{\rm PS}/m_{\rm V})}{\partial \beta} \\
 \frac{\partial (m_{\rm V}a)}{\partial K} &
 \frac{\partial (m_{\rm PS}/m_{\rm V})}{\partial K} 
\end{array} \right)^{-1}.
\label{eq:massmat}
\end{eqnarray} 
In practice, however, we find that there exists a region where the matrix 
on the right hand side becomes almost singular, so that the inverse cannot 
be solved reliably.
In particular, when quarks are heavy, because $m_{\rm PS}/m_{\rm V}$ 
is always close to one, its derivatives in terms of $\beta$ and $K$ 
cannot be determined precisely. 

This leads us to adopt the following alternative method.
We determine 
$m_{\rm V} a \frac{\partial \beta}{\partial m_{\rm V} a}$ and 
$m_{\rm V} a \frac{\partial K}{\partial m_{\rm V} a}$ 
directly from the inverse functions 
$\beta(m_{\rm V} a, m_{\rm PS}/m_{\rm V})$ and 
$K(m_{\rm V} a, m_{\rm PS}/m_{\rm V})$. 
In Fig.~\ref{fig:pirho},
we plot results for $(m_{\rm V} a, m_{\rm PS}/m_{\rm V})$ 
at zero temperature. 
To determine 
$a \frac{\partial \beta}{\partial a}$ and $a \frac{\partial K}{\partial a}$ 
at a point, say, $((m_{\rm V} a)_0, (m_{\rm PS}/m_{\rm V})_0)$, 
we fit data for $(\beta, K)$, {\it i.e.,} the values specifying the 
simulation points, by a power function expanded in terms of 
$(\Delta (m_{\rm V} a), \Delta (m_{\rm PS}/m_{\rm V})) 
= (m_{\rm V} a - (m_{\rm V} a)_0, (m_{\rm PS}/m_{\rm V}) - 
(m_{\rm PS}/m_{\rm V})_0)$.  
We employ the following general fit ansatz up to the third power, 
\begin{eqnarray}
\beta &=& c_{\beta 0} + c_{\beta 1} \{\Delta m_{\rm V}a \}
          + c_{\beta 2} \{\Delta m_{\rm V}a \}^2
          + c_{\beta 3} \{\Delta m_{\rm V}a \}^3
          + c_{\beta 4} \{\Delta (m_{\rm PS}/m_{\rm V})\}  \nonumber \\
      & & + c_{\beta 5} \{\Delta (m_{\rm PS}/m_{\rm V})\}
                \{\Delta m_{\rm V}a\} 
          + c_{\beta 6} \{\Delta (m_{\rm PS}/m_{\rm V}) \}
                \{\Delta m_{\rm V}a \}^{2}
          + c_{\beta 7} \{\Delta (m_{\rm PS}/m_{\rm V}) \}^{2}
                                               \nonumber \\
      & & + c_{\beta 8} \{\Delta (m_{\rm PS}/m_{\rm V}) \}^{2}
                \{m_{\rm V}a - (m_{\rm V}a)_0\}
          + c_{\beta 9} \{\Delta (m_{\rm PS}/m_{\rm V}) \}^{3}
                    \label{eq:betafit} \\
K &=& c_{K 0} + c_{K 1} \{\Delta m_{\rm V}a \}
          + c_{K 2} \{\Delta m_{\rm V}a \}^2
          + c_{K 3} \{\Delta m_{\rm V}a \}^3
          + c_{K 4} \{\Delta (m_{\rm PS}/m_{\rm V}) \}  \nonumber \\
      & & + c_{K 5} \{\Delta (m_{\rm PS}/m_{\rm V}) \}
                    \{\Delta m_{\rm PS}a \} 
          + c_{K 6} \{\Delta (m_{\rm PS}/m_{\rm V}) \}
                \{\Delta m_{\rm V}a \}^{2}
          + c_{K 7} \{\Delta (m_{\rm PS}/m_{\rm V}) \}^{2} \nonumber \\
      & & + c_{K 8} \{\Delta (m_{\rm PS}/m_{\rm V}) \}^{2}
                \{\Delta m_{\rm V}a \}
          + c_{K 9} \{\Delta (m_{\rm PS}/m_{\rm V}) \}^{3}. \label{eq:Kfit} 
\end{eqnarray}
The fit range is determined for each 
$((m_{\rm V} a)_0, (m_{\rm PS}/m_{\rm V})_0)$ separately: 
The fit range in $m_{\rm PS}/m_{\rm V}$ is fixed such that  
$\chi^2/N_{\rm df}$ is minimized, under the condition that 
the number of fitted data is larger than 30 to avoid artifacts 
from statistical fluctuations.
For the fit range in $m_{\rm V}a$, we include all data 
except for the points $m_{\rm V}a < 2.3$ or $K \geq 0.11$, 
which are far from the region we study.

From the fit results, we calculate the beta functions by 
differentiating $\beta$ and $K$ in terms of $m_{\rm V} a$, 
with $m_{\rm PS}/m_{\rm V}$ fixed, 
\begin{eqnarray}
m_{\rm V}a \frac{\partial \beta}{\partial (m_{\rm V}a)}
&=& m_{\rm V}a \, c_{\beta 1}, \\
m_{\rm V}a \frac{\partial K}{\partial (m_{\rm V}a)}
&=& m_{\rm V}a \, c_{K 1}.
\end{eqnarray}
The results are plotted in Figs.~\ref{fig:dbdrho} and \ref{fig:dkdrho}. 
In Fig.~\ref{fig:dbdrho}, the one-loop perturbative value of 
$m_{\rm V}a \frac{\partial \beta}{\partial (m_{\rm V}a)}$ for 
massless quark is shown by a solid line near the right edge of the plot. 
Our results appear to gradually approach this value in the large 
$\beta$ limit. 
We also see that $m_{\rm V}a \frac{\partial K}{\partial (m_{\rm V}a)}$ 
for small $m_{\rm PS}/m_{\rm V}$ approaches zero at large $\beta$, 
as we expect from the fact that $K_c \rightarrow 1/8$ 
as $\beta \rightarrow \infty$.

Another application of the fits (\ref{eq:betafit}) and (\ref{eq:Kfit}) is  
the determination of the lines of constant physics and the 
temperature measured by the vector meson mass, 
$T/m_{\rm V} = 1/(N_t m_{\rm V} a)$, 
discussed in the previous section.

\section{Equation of state}
\label{sec:eos}

The energy density and pressure are defined by
\begin{eqnarray}
\epsilon = - \frac{1}{V} \frac{\partial \ln Z}{\partial T^{-1}}, 
\hspace{10mm} 
 p       = T \frac{\partial \ln Z}{\partial V},
\end{eqnarray}
where $Z$, $T$ and $V$ are the partition function, temperature 
and spatial volume, respectively.  We calculate these quantities 
as a function of temperature along the lines of constant physics 
obtained in Sec.~\ref{sec:scale}.

\subsection{Integral method in full QCD}
\label{sec:method}

We compute the pressure by the integral method \cite{Eng90}. 
This method is based on the formula $p= -f$, with 
$f = (-T/V) \ln Z$ the free energy density, 
valid for large homogeneous systems.
The pressure is then given by 
\begin{eqnarray} 
\label{eq:integral}
\frac{p}{T^{4}} = -\frac{f}{T^{4}} = - N_{t}^{4} 
\int^{(\beta,K)} {\rm d} \xi \left\{ 
\left\langle \frac{1}{N_{s}^{3} N_{t}} 
      \frac{\partial S}{\partial \xi} (\beta',K') \right\rangle_{\rm sub}
\right\}
\end{eqnarray}
where 
${\rm d}\xi=({\rm d}\beta',{\rm d}K')$
is the line element in the $(\beta,K)$ plane, 
and $\left\langle \cdots \right\rangle_{\rm sub}$ is the expectation value 
at finite temperature with the zero-temperature value subtracted. 
The starting point of the integration path should be chosen 
in the low temperature phase where $p\approx 0$.
In actual simulations, for setting the starting point of the integration path,
we quadratically extrapolate results for the integrand near the low temperature 
phase to zero. 

For our action (\ref{eq:gaction}) and (\ref{eq:qaction}), 
the derivatives in (\ref{eq:integral}) are given by 
\begin{eqnarray}
\frac{\partial {\rm ln} Z}{\partial \beta} 
&=& \left\langle -\frac{\partial S}{\partial \beta} \right\rangle 
= N_{s}^{3} N_{t}
\left( \left\langle c_0 \sum_{x, \mu > \nu} W^{1 \times 1}_{\mu \nu} 
     + c_1 \sum_{x,\mu,\nu} W^{1 \times 2}_{\mu \nu}
             \right\rangle \right. \nonumber \\
& & \hspace{3cm} \left. - N_{f} \frac{\partial c_{SW}}{\partial \beta} K 
   \left\langle \sum_{x, \mu > \nu} {\rm Tr}^{(c,s)}
    \sigma_{\mu \nu} F_{\mu \nu}(x) 
    D^{-1}(x,x) \right\rangle \right) \\ \nonumber \\
    \frac{\partial {\rm ln} Z}{\partial K} 
&=& \left\langle -\frac{\partial S}{\partial K} \right\rangle 
= N_{f} N_{s}^{3} N_{t} 
\left( \left\langle -  \sum_{x, \mu} 
 {\rm Tr}^{(c,s)} \{ (1-\gamma_{\mu})U_{\mu}(x) 
 D^{-1}(x+{\hat\mu},x) \right. \right. \hspace{20mm} \nonumber \\ 
 & &\left. \left.  
 + (1+\gamma_{\mu})U^{\dagger}_{\mu}(x) D^{-1}(x,x+{\hat\mu}) \} \right\rangle 
 - c_{SW} \left\langle \sum_{x, \mu > \nu} {\rm Tr}^{(c,s)} 
 \sigma_{\mu \nu} F_{\mu \nu}(x) D^{-1}(x,x) \right\rangle \right),
\end{eqnarray}
where $N_s$ is the spatial lattice size and $N_f=2$ denotes the 
number of flavors.  
We evaluate the quark contributions, 
$\frac{\partial S}{\partial \beta}$ and 
$\frac{\partial S}{\partial K}$, using the noisy source method. 
In order to select the type of noise, we have tested $Z(2)$, $U(1)$ 
and a complex Gaussian noise with a test run on an $8^3 \times 4$ lattice. 
We find that the $U(1)$ and $Z(2)$ noises show faster convergence 
than the Gaussian noise in the number of noise ensembles. 
The difference between the $U(1)$ and $Z(2)$ cases is small, while 
the $U(1)$ noise shows slightly faster convergence in this test. 
From this result, we have adopted the $U(1)$ noise in this study.

The integral method was originally developed for a pure gauge system, 
for which the parameter space is one-dimensional. 
In our case, the parameter space is two-dimensional. 
Therefore, the integration path for the pressure (\ref{eq:integral}) is 
not unique.  
We have performed a series of test runs on an $8^3 \times 4$ lattice, 
and have confirmed the integration path independence \cite{eoslat99}.
Details of the test are presented in Appendix~\ref{sec:app1}.  
We shall present results of a similar test in our production runs below. 

We also find from this test that 
the integration paths in the $K$ direction (constant $\beta$) 
lead to smaller errors for the final values of the pressure 
than those in the $\beta$ direction (constant $K$). 
Therefore, we choose the integration paths in the $K$ direction starting 
from the region of small values of $K$ and moving towards the chiral limit.
Our simulation points on the $16^3\times4$ and $16^3\times6$ lattices are 
shown by ``$+$'' and ``$\times$'' in Fig.~\ref{fig:phase}.

\subsection{Equation of state for $N_t=4$}
\label{sec:eosnt4}

In Fig.~\ref{fig:dpdk}, we show the results for the pressure derivative 
\begin{eqnarray}
\frac{\partial (p/T^4)}{\partial K} = - N_t^4 
\left\langle \frac{1}{N_{s}^{3} N_{t}}
\frac{\partial S}{\partial K} \right\rangle_{\rm sub}
\end{eqnarray}
obtained on an $N_t=4$ lattice. 
Measurements are made with five noise ensembles at every trajectory. 
The bin size for the jack-knife errors is set to 10 trajectories 
from a study of bin size dependence. 
Numerical values for the derivative are summarized in Table~\ref{tab:dsdk}.
Interpolating the data by a cubic spline method, we integrate in the $K$ 
direction to obtain the pressure presented in Fig.~\ref{fig:prsk}.

We also compute the derivative in the $\beta$ direction, 
\begin{eqnarray}
\frac{\partial (p/T^4)}{\partial \beta} = - N_t^4 
\left\langle \frac{1}{N_{s}^{3} N_{t}}
\frac{\partial S}{\partial \beta} \right\rangle_{\rm sub},
\label{eq:dpdb}
\end{eqnarray}
as shown in Fig.~\ref{fig:dpdb} and Table~\ref{tab:dsdb}.
We observe that the results for this derivative are noisier than those for 
the $K$ derivative in Fig.~\ref{fig:dpdk}.
This is the underlying reason for the fact commented in Sec.~\ref{sec:method}
that the integral paths in the $K$ direction lead to smaller errors in 
the pressure.

In Fig.~\ref{fig:prsb}, we replot the pressure from the $K$ integration 
as a function of $\beta$, and compare it with the slope 
(\ref{eq:dpdb}) computed independently from the 
$\frac{\partial S}{\partial \beta}$ data.  
The latter data for the slope, shown by short lines,  
are tangential to the pressure curve, 
confirming the integration path independence of results for the pressure.

The pressure data shown in Fig.~\ref{fig:prsk} or \ref{fig:prsb} are 
not quite useful yet.  
We wish to compute the pressure on a line of constant physics 
as a function of temperature normalized by 
the pseudo-critical temperature $T_{pc}$ on the same line of constant 
physics. 
The necessary change of parameters from $(\beta, K)$ to 
$(m_{\rm PS}/m_{\rm V}, T/T_{pc})$ is achieved with the interpolations 
performed in Secs.~\ref{sec:scale} and \ref{sec:betafnc}.

The pressure $p/T^4$ as a function of $T/T_{pc}$ is given in 
Fig.~\ref{fig:prsT4} for $m_{\rm PS}/m_{\rm V} = 0.65$, 0.7, 0.75, 0.8, 
0.85, 0.9 and 0.95.
In this figure, symbols denote the values on the integration path along the 
$K$ direction at $\beta=1.80$, 1.85, 1.90, 1.95, 2.00, 2.10 and 2.20,  
{\it i.e.,} for each $\beta$, the pressure in Fig.~\ref{fig:prsk} 
at the values of $K$ corresponding to the given values of 
$m_{\rm PS}/m_{\rm V}$. 
The values of $T/T_{pc}$ at those points are determined from 
$m_{\rm V}a$, as discussed in Sec.~\ref{sec:scale}.
To interpolate these symbols in the direction of $\beta$ ({\it i.e.,} of 
$T/T_{pc}$ ), 
we use the results for the slopes $\frac{\partial S}{\partial \beta}$ 
shown in Fig.~\ref{fig:prsb} and adopt a cubic ansatz.

We observe in Fig.~\ref{fig:prsT4} that the pressure depends only 
weakly on the quark mass once the ratio $m_{\rm PS}/m_{\rm V}$ falls 
below $\approx 0.8$.  In the heavy quark limit $m_{\rm PS}/m_{\rm V}=1$, 
the pressure should coincide with the pure gauge value on a lattice 
with the same size, which is denoted by a dashed line \cite{okamoto}. 
While the pressure decreases for $m_{\rm PS}/m_{\rm V}=0.80$--0.95,  
the values at $m_{\rm PS}/m_{\rm V}=0.95$ are still large compared to 
those of the pure gauge system for $N_t=4$.

In Fig.~\ref{fig:prsT4}, we also mark the high-temperature 
Stefan-Boltzmann (SB) values by solid horizontal lines, both for 
$N_t=4$ and in the continuum. 
The lattice value is evaluated from the free energy density in the SB limit, 
to be in parallel with the integral method adopted in numerical simulations. 
Some details of this computation are described in Appendix~\ref{sec:app2}.
We observe that the pressure overshoots the SB value in the continuum limit, 
and appears to gradually increase toward the SB value for the $N_t=4$ lattice
at high temperatures.
Another point to note is that the large SB value 
on an $N_t=4$ lattice \cite{okamoto,Eng82b} is dominated by the quark 
contribution. 
While the integral method does not allow a separate evaluation of the 
two contributions, a comparison of the two-flavor result and that 
of the pure gluon theory \cite{okamoto} (dashed line) 
shows that the situation should be similar at finite temperatures.    

To compute the energy density, we use the following expression for 
the interaction measure $\epsilon - 3p$:
\begin{eqnarray}
\frac{\epsilon - 3p}{T^4} 
&=& N_t^4 \left\langle \frac{1}{N_s^3 N_t} 
a \frac{\partial S}{\partial a} \right\rangle_{\rm sub} \nonumber \\
&=& N_t^4 \left[
a \frac{\partial \beta}{\partial a} 
 \left\langle \frac{1}{N_s^3 N_t} 
 \frac{\partial S}{\partial \beta} \right\rangle_{\rm sub}
+ a \frac{\partial K}{\partial a} 
 \left\langle  \frac{1}{N_s^3 N_t} 
 \frac{\partial S}{\partial K} \right\rangle_{\rm sub} \right].
\label{eq:e3p}
\end{eqnarray}
Applying the beta functions calculated in Sec.~\ref{sec:betafnc}, 
we find the results for $(\epsilon - 3p)/T^4$ shown in Fig.~\ref{fig:e3pT4}.
The meaning of symbols is the same as in Fig.~\ref{fig:prsT4}. 

Combining Figs.~\ref{fig:prsT4} and \ref{fig:e3pT4} for $p/T^4$ and 
$(\epsilon - 3p)/T^4$, we obtain the energy density 
plotted in Fig~\ref{fig:engT4}. 
This quantity also overshoots the SB value in the continuum limit. 
In contrast to the case of pressure, the energy density in the high 
temperature phase is quite constant as a function of temperature.  

Our results for pressure and energy density 
allow us to calculate the speed of sound $c_s$ 
defined by
\begin{equation}
c_s^2 = \frac{\partial p}{\partial \epsilon}. 
\end{equation}
We compute the derivative from a quadratic fit of $p$ as a function of 
$\epsilon$ using three data points along the lines of constant physics. 
The results for $N_t=4$ are plotted in Fig.~\ref{fig:souT} where errors are 
estimated by error propagation from those of $p$ and $\epsilon$.  
We omit results at small temperatures with $T/T_{ps} < 0.9$, 
since there the magnitude of $\epsilon$ is small in comparison with its error.
The speed of sound rapidly increases just 
above the critical temperature, and almost saturates the SB value 
when $T/T_{pc} \simgt 1.5$.

\subsection{Equation of state for $N_t=6$}
\label{sec:eosnt6}

The simulation points for $N_t=6$ lattices at $\beta = 1.95$, 2.00, 2.10, 
2.20, and 2.30 are marked by crosses in Fig.~\ref{fig:phase}.
Raw data for the derivatives $\partial (p/T^4)/ \partial K$ and 
$\partial (p/T^4)/\partial \beta$ are shown in 
Fig.~\ref{fig:dpdk6} and Fig.~\ref{fig:dpdb6}, respectively.
Since the statistical errors for $N_t=6$ results are larger than 
those for $N_t=4$, 
the spline interpolation does not work well for $N_t=6$.
Therefore, we interpolate the pressure derivatives by straight lines.
The rest of data analyses parallel those for the case of $N_t=4$. 

The pressure for $N_t=6$ on the lines of constant physics 
are plotted in Fig.~\ref{fig:prsT6} by open symbols 
as a function of $T/T_{pc}$, 
together with the results for $N_t=4$ (filled symbols). 
Figure~\ref{fig:e3pT6} shows the interaction measure 
$(\epsilon - 3p)/T^4$ as a function of $T/T_{pc}$, calculated from results 
in Figs.~\ref{fig:dpdk6} and \ref{fig:dpdb6} together with the beta functions
of Figs.~\ref{fig:dbdrho} and \ref{fig:dkdrho}. 
Combining the results for pressure and interaction measure, we obtain 
the energy density presented in Fig.~\ref{fig:engT6}. 

These figures show that both the pressure and energy density decreases 
as the lattice spacing becomes smaller from $N_t=4$ to 6, and the values 
at high temperatures become closer to the continuum SB limit. 
On the other hand, both at $N_t=4$ and 6, the energy density is smaller 
than the SB values for the corresponding $N_t$, and an approach to the 
lattice SB value toward high temperatures is not apparent in our 
data.  A similar deviation of energy density from the SB value at finite 
$N_t$ has been reported in Ref.~\cite{okamoto} for the case of the SU(3) 
pure gauge theory using the RG improved gauge action (\ref{eq:gaction}). 
Further study is necessary to examine if deviations remain toward 
the limit of high temperatures. 

We also observe for both the pressure and the energy density that 
the dependence on the quark mass is quite small for 
$m_{\rm PS}/m_{\rm V} \approx 0.65$--0.8. 
A weak quark mass dependence appears only at 
$m_{\rm PS}/m_{\rm V} \simgt 0.9$ for $N_t=4$ (the errors are still 
too large to conclude a quark mass dependence for the $N_t=6$ data).
This result may not be surprising since hadron mass results in our 
zero-temperature simulations 
\cite{CPPACSmq} show that the renormalized quark mass at $\mu=2$~GeV 
in the $\overline{MS}$ scheme at $m_{\rm PS}/m_{\rm V}\approx 0.65$--0.8 
equals 
$m_q^{\overline{MS}}(\mu=2 \mbox{GeV})\approx 100$--200~MeV, which is 
similar in magnitude to the critical temperature $T_c\approx 170$~MeV 
estimated for two-flavor QCD \cite{ejiri00}.  
For comparison, finite mass corrections 
for free fermion gas only amount to 7\% when the temperature equals 
the fermion mass $m_f$, and exceed 50\% only when $m_f/T \simgt 3$.

In a previous study using the standard staggered quark action, it was reported 
that the energy density $\epsilon/T^4$ for $N_t=4$ overshoot the SB value 
forming a sharp peak just above $T_c$, 
while the results for $N_t=6$ show no peak \cite{Ber97b}. 
With the improved Wilson quark action, we do not observe an overshoot 
of the energy density both at $N_t=4$ and 6. 
Similar absence of the peak of energy density is reported also from 
an improved staggered quark action at $N_t=4$ when a contribution 
proportional to the bare quark mass is removed \cite{Kar00,Kar01}.
We think it likely that the overshoot observed with staggered quark action at 
$N_t=4$ is a lattice artifact.
With the staggered quark action, the energy densities for $m_q/T=0.075$ 
and 0.15 at $N_t=6$ are found to be consistent with each other within 
the errors \cite{Ber97b}. 
This result is similar to our finding of small quark mass dependence in 
the EOS. 

Our present data for $N_t=4$ and 6 show a 50\% decrease both in the 
pressure and energy density, which is too large to attempt a continuum 
extrapolation.  On $N_t=6$ lattices, however, the magnitude and temperature 
dependence of the two quantities are quite similar 
between our improved Wilson quark action and the staggered quark action. 
Together with the fact that the $N_t=6$ results are close to the continuum 
SB limit at high temperatures, the approximate agreement of EOS between 
two different types of actions may be suggesting that 
the $N_t=6$ results are not far from the continuum limit. 
This expectation is also supported by the $N_t$ dependence of the SB value 
on the lattice.  The value for $N_t=6$, which is 50\% too large compared 
to the continuum limit, reduces by 30\% so that for $N_t=8$ the lattice 
SB value is within 20\% of the continuum limit. 
Thus we expect that a precise continuum extrapolation will be possible if 
additional data points at $N_t=8$ are generated, which we leave for future 
work.

\section{Conclusion}
\label{sec:concl}

We have presented first results for the equation of state in QCD 
with two flavors of dynamical quarks using a Wilson-type quark action. 
In order to suppress large lattice artifacts observed with the standard 
Wilson quark action, we have adopted a clover-improved form of the 
action and an RG-improved gluon action.  
Two temporal lattice sizes, $N_t=4$ and 6, are studied to examine 
the magnitude of finite lattice spacing errors.

We have calculated the energy density and the pressure as functions 
of temperature along the lines of constant physics, which are identified 
through the mass ratio $m_{\rm PS}/m_{\rm V}$.  As a part of the analysis 
to work out these lines, we have also computed the beta functions in the 
parameter space $(\beta,K)$. 

We found that the quark mass dependence of EOS is small over the range 
$m_{\rm PS}/m_{\rm V} \approx 0.65$--0.8. 
While the physical point $m_{\rm PS}/m_{\rm V}=0.18$ is still far away,  
the observed independence on the quark mass suggests 
that our result for the EOS is close to those at the physical point
except in the vicinity of the chiral transition point 
where a singular limit according to the O(4) critical
exponents is expected.

Our results for the pseudo-critical temperature in unit of the vector 
meson mass, $T_{pc}/m_{\rm V}$, show an agreement within about 10\% between 
the temporal lattice sizes $N_t=4$ and 6.
On the other hand, the pressure and the energy density decreases 
substantially, showing the presence of large scaling violation 
in the results for $N_t=4$. 
An encouraging indication, however, is that results on the $N_t=6$ lattice 
are close to the continuum Stefan-Boltzmann value at high temperatures. 
We also note that the values and the temperature dependence of EOS at 
$N_t=6$ are quite similar to the previous results from staggered quark 
action \cite{Ber97b}. 
These may be suggestions of a possibility that precise calculations of  
EOS are realized on lattices with temporal sizes $N_t$ not 
much larger than 6.

\section*{Acknowledgements}

This work is supported in part by Grants-in-Aid of the Ministry of 
Education (Nos.~10640246, 10640248, 11640250, 11640294, 
11740162, 12014202, 12304011, 12640253, 12740133). 
AAK and TM are supported by the Research for Future Program of JSPS
(No. JSPS-RFTF 97P01102). 
SE, JN, KN, M. Okamoto and HPS are JSPS Research Fellows. 


\input appendix.tex


\newpage
\input tables.tex

\newpage
\input figures.tex


\end{document}

%% file: appendix.tex
\appendix

\section{A test of the integral method in full QCD}
\label{sec:app1}

We compute EOS by the integral method \cite{Eng90} described in 
Sec.~\ref{sec:method}.
In order to test the method, we perform a series of test runs 
to calculate the pressure on an $8^3 \times 4$ lattice. 
For subtraction of the zero temperature part, we also measure the same 
operators on an $8^4$ lattice. 
Simulation points are shown by stars in Fig.~\ref{fig:path84}.
We generate 500 HMC trajectories at each point.

We first check the influence of the spatial volume on EOS. 
In order to avoid systematic errors from numerical interpolation and 
extrapolation needed for numerical integrations, 
we first compare the values of the integrand $\partial (p/T^4)/\partial K$ 
for $N_s/N_t=2$ ($8^3 \times 4$ lattice) with those for $N_s/N_t=4$ 
from our main runs on a $16^3\times4$ lattice.
From Fig.~\ref{fig:voldep_dpdk},
we find that the diffirence between $N_s/N_t=2$ and 4 is comparable 
with statistical fluctuations. 
Integrating out these values, we obtain Fig.~\ref{fig:voldep_p}. 
We find that the two results agree well with each other --- 
the slight discrepancy of the $\beta=2.2$ data at small $K$ seems to be 
caused by a longer extrapolation to the point 
$\partial (p/T^4)/\partial K = 0$ for the stating point of the integration.

We then test the integration path independence in the integral method.
We study three integration paths in the parameter space $(\beta,K)$, 
shown in Fig.~\ref{fig:path84}. 
At the crossing points, the results for the pressure from different 
paths should coincide with each other.
The results for $p/T^4$ obtained from these paths are summarized in 
Fig.~\ref{fig:prs84}.
The left figure is obtained by integrating in the $\beta$ direction 
at $K=0.13$, 
while the right figure is computed from the integration paths 
in the $K$ direction at $\beta = 2.1$ and 2.2. 

We find that $p/T^4$ at $(\beta, K) = (2.1,0.13)$ and $(2.2,0.13)$ 
in the two figures agree well with each other. 
This confirms the path-independence of the pressure. 

We also note that the paths in the $K$ direction lead to much 
smaller errors in the pressure than the path in the $\beta$ direction. 
We therefore adopt paths in the $K$ direction in the production runs 
discussed in the main text. 

\section{Stefan-Boltzmann limit of pressure by the integral method}
\label{sec:app2}

In the calculation of pressure discussed in Sec.~\ref{sec:eos}, 
we employ the integral method, in which the negative of free energy density 
$-f=(T/V){\rm ln}Z$ is identified with the pressure. 
In order to compute the Stefan-Boltzman value to be compared 
with the pressure from the integral method, we should compute the 
free energy density in the free gas limit. 
In this appendix, we describe our calculation of the free energy density 
for the case of our improved lattice action.

To calculate the partition function $Z$, 
we expand the link variable as
\begin{eqnarray}
U_{\mu} (x) = \exp \{ig A_{\mu}(x) \} = \exp \{ig A_{\mu}^a (x) T_a \}
\end{eqnarray}
and perform a Fourier transformation 
\begin{eqnarray}
A_{\mu}^a (x) = \frac{1}{\sqrt{N_s^3 N_t}} \sum_{k}
{\rm e}^{ik (x + \hat{\mu} /2)} A_{\mu}^a (k),
\end{eqnarray}
where 
\begin{eqnarray}
k_{\mu} &=& \frac{2 \pi j_{\mu}}{N_s}, \ \
j_{\mu}=0, \pm1, \cdots, N_s/2 \ \ {\rm for \ \mu=1,2,3} \\
k_{\mu} &=& \frac{2 \pi j_{\mu}}{N_t}, \ \
j_{\mu}=0, \pm1, \cdots, N_t/2 \ \ {\rm for \ \mu=4}.
\end{eqnarray}
Fixing the gauge to the lattice Lorentz gauge 
by adding the gauge fixing term
\begin{eqnarray}
\label{eq:gaugefix}
 S_{\rm gf} = - \sum_{x} {\rm Tr}
\left( \sum_{\mu} \nabla_{\mu} A_{\mu} (x) \right)^2,
\end{eqnarray}
the free part of the gauge action (\ref{eq:gaction}) is given by 
\begin{eqnarray}
S_{g}^{(0)} + S_{\rm gf}^{(0)} = -\frac{1}{4} \sum_{k, \mu, \nu, a}
[q_{\mu \nu} (k) \hat{F}_{\mu \nu}^a (k) \hat{F}_{\mu \nu}^a (-k)
+ 2 \hat{k}_{\mu} \hat{k}_{\nu} A_{\mu}^a (k) A_{\nu}^a (-k)],
\end{eqnarray}
where 
$\hat{k}_{\mu}= 2 \sin (k_{\mu} / 2)$,
$\hat{F}_{\mu \nu}^a (k) =
i(\hat{k}_{\mu} A_{\nu}^a (k) - \hat{k}_{\nu} A_{\mu}^a (k))$,
and 
\begin{eqnarray}
q_{\mu \nu} &=& 1 - c_1 (\hat{k}_{\mu}^2 + \hat{k}_{\nu}^2); \ \ 
{\rm for \ \mu \neq \nu} \\
q_{\mu \nu} &=& 0; \hspace{15mm} {\rm for \ \mu = \nu}.
\end{eqnarray}
The free part of the ghost term corresponding to 
the gauge fixing (\ref{eq:gaugefix}) is given by 
\begin{eqnarray}
 S_{\rm gh}^{(0)} = \sum_{k, \mu, a} \hat{k}_{\mu} \hat{k}_{\mu}
\bar{\eta}^a (-k) \eta^a (k),
\end{eqnarray}
where $\eta$ and $\bar{\eta}$ are the ghost fields. 

The partition function for the gauge part can be calculated as 
\begin{eqnarray}
  Z_{g}= \left[ \prod_{k}
  \det D_{\mu \nu}^{1/2} (k) D_{\rm gh}^{-1} (k) \times {\rm (const.)}
  \right]^{8},
\end{eqnarray}
with
\begin{eqnarray}
D_{\mu \nu}^{-1} (k) &=&  \hat{k}_{\mu} \hat{k}_{\nu} + 
\sum_{\rho}(\hat{k}_{\rho} \delta_{\mu \nu} - \hat{k}_{\mu} \delta_{\rho \nu}) 
q_{\mu \rho} \hat{k}_{\rho}, \\
D_{\rm gh}^{-1} (k) &=& \sum_{\mu} \hat{k}_{\mu}^2.
\end{eqnarray}
Consequently, we obtain the gauge part of the unnormalized free energy density
\begin{eqnarray}
  f^{(g)} a^4 = -\frac{1}{N_s^3 N_t} \ln Z_g =
  \frac{8}{N_s^3 N_t} {\sum_{k}}' \left[ \frac{1}{2}
  \ln(\det D_{\mu \nu}^{-1}(k))
  - \ln D_{\rm gh}^{-1}(k) \right] + {\rm (const.)}.
\end{eqnarray} 
where $\sum'$ means a sum except for the zero mode.

The free part of the quark action (\ref{eq:qaction}) is given by 
\begin{eqnarray}
S_q & = & - N_f \sum_{a=1}^{3} \sum_{k}\overline{q}^a (k) D_{q} q^a (k),\\
D_{q}(k) & = & 2K \sum_{\mu=1}^{4} \cos(k_{\mu}) -1
-2iK \sum_{\mu =1}^{4} \gamma_{\mu} \sin(k_{\mu}),
\end{eqnarray}
where
\begin{eqnarray}
k_{\mu} &= \frac{2 \pi j_{\mu}}{N_s}, \ \
j_{\mu}=0, \pm1, \cdots, N_s/2 \ \ & {\rm for \ \mu=1,2,3} \\
k_{\mu} &= \frac{2 \pi (j_{\mu}+ 1/2)}{N_t}, \ \
j_{\mu}=0, \pm1, \cdots, N_t/2 \ \ & {\rm for \ \mu=4}.
\end{eqnarray}
The partition function for the quark part is obtained as 
\begin{eqnarray}
Z_{q} = \left( \prod_{k} \det D_{q}(k) \right)^{3 N_f}.
\end{eqnarray}
\begin{eqnarray}
\det D_{q}(k) = 
\left[ \left\{1-8K+4K \sum_{\mu=1}^4 \sin^2(k_{\mu}/2) \right\}^2
+4K^2 \sum_{\mu=1}^4 \sin^2 (k_{\mu}) \right]^2 .
\end{eqnarray}
We then obtain the quark part of the unnormalized free energy 
density \cite{Eng82b}.
\begin{eqnarray}
  f^{(q)} a^4 = -\frac{1}{N_s^3 N_t} \ln Z_q =
  - \frac{3 N_f}{N_s^3 N_t} \sum_{k} \ln( \det D_{q}(k)).
\end{eqnarray}

Numerical calculations of the normalized energy density and 
pressure are performed using the equations, 
\begin{eqnarray}
p a^4 &=& -(f^{(g)} - f_{T=0}^{(g)}) a^4 -(f^{(q)} - f_{T=0}^{(q)}) a^4 , \\
\epsilon a^4 &=& 3 p a^4,
\end{eqnarray}
where $f_{T=0}^{(g)}$ and $f_{T=0}^{(q)}$ are the free energy at zero 
temperature calculated on an $N_s^4$ lattice.
At the right edge of the figures \ref{fig:prsT4}, \ref{fig:engT4}, 
\ref{fig:prsT6} and \ref{fig:engT6}, we show the results 
for the cases of $16^3 \times 4$ and $16^3 \times 6$ lattices, 
in the massless quark limit, $K=1/8$.

%% file: tables.tex
\begin{table} [htb]
\begin{center}
  \begin{tabular}{cccc}
$\beta$ & $K$ & traj. & therm. \\ \hline
1.80 & 0.1300--0.1450 &  500--2000 & 200--500 \\ 
1.85 & 0.1250--0.1440 &  500--1900 & 200--300 \\ 
1.90 & 0.1250--0.1425 &  500--2000 & 200--400 \\ 
1.95 & 0.1200--0.1410 &  500--2000 & 200 \\ 
2.00 & 0.1150--0.1390 &  500--2000 & 200--300 \\ 
2.10 & 0.0900--0.1375 &  500--1000 & 200--900 \\ 
2.20 & 0.0700--0.1365 &  500 & 200 \\ 
\end{tabular}
\end{center}
\caption{Simulation parameters on a $16^{3} \times 4$ lattice.}
\label{tab:param16x4}
\end{table}

\begin{table} [htb]
\begin{center}
  \begin{tabular}{cccc}
$\beta$ & $K$ & traj. & therm. \\ \hline
1.95 & 0.1350--0.1410 &  1000 & 200 \\ 
2.00 & 0.1300--0.1395 &  800--1500 & 200 \\ 
2.10 & 0.1200--0.1375 &  500--1000 & 200--300 \\ 
2.20 & 0.1100--0.1365 &  500--1000 & 200--300 \\ 
2.30 & 0.1000--0.1360 &  500--1500 & 200--250 \\ 
\end{tabular}
\end{center}
\caption{Simulation parameters on a $16^{3} \times 6$ lattice.}
\label{tab:param16x6}
\end{table}

\begin{table} [htb]
\begin{center}
  \begin{tabular}{cccc}
$\beta$ & $K$ & traj. & therm. \\ \hline
1.80 & 0.1300--0.1450 & 200 & 200--500 \\ 
1.85 & 0.1250--0.1440 & 200--300 & 100--300 \\ 
1.90 & 0.1250--0.1425 & 200 & 200--400 \\ 
1.95 & 0.1200--0.1410 & 200--300 & 100--400 \\ 
2.00 & 0.1150--0.1395 & 200--300 & 100--200 \\ 
2.10 & 0.0900--0.1375 & 300 & 200--550 \\ 
2.20 & 0.0700--0.1365 & 200--300 & 100--200 \\ 
2.30 & 0.1000--0.1360 & 200 & 100 \\ 
\end{tabular}
\end{center}
\caption{Simulation parameters on a $16^{4}$ lattice.}
\label{tab:param16x16}
\end{table}

\begin{table}[hbt]
\begin{center}
  \begin{tabular}{lllll}
$\beta$ & $K$ & $m_{\rm PS}a$ & $m_{\rm V}a$ & $m_{\rm PS}/m_{\rm V}$ \\ 
\hline
1.80 & 0.1300 & 1.7677(41) & 1.9318(47) & 0.9150(14) \\
1.80 & 0.1350 & 1.5329(40) & 1.7384(49) & 0.8818(14) \\
1.80 & 0.1375 & 1.3883(45) & 1.6310(62) & 0.8512(19) \\
1.80 & 0.1400 & 1.2094(35) & 1.4769(51) & 0.8189(25) \\
1.80 & 0.1425 & 1.0222(44) & 1.3368(78) & 0.7646(33) \\
1.80 & 0.1440 & 0.8783(63) & 1.2198(76) & 0.7200(46) \\
1.80 & 0.1450 & 0.7569(46) & 1.1563(88) & 0.6546(54) \\
\hline
1.85 & 0.1250 & 1.9104(48) & 2.0410(52) & 0.9360(11) \\
1.85 & 0.1300 & 1.6816(35) & 1.8440(43) & 0.9120(11) \\
1.85 & 0.1350 & 1.4107(44) & 1.6148(52) & 0.8736(21) \\
1.85 & 0.1375 & 1.2531(32) & 1.4862(46) & 0.8431(16) \\
1.85 & 0.1400 & 1.0463(38) & 1.3170(70) & 0.7945(28) \\
1.85 & 0.1425 & 0.8054(37) & 1.1233(67) & 0.7169(33) \\
1.85 & 0.1440 & 0.5635(46) & 0.9488(71) & 0.5939(50) \\
\hline 
1.90 & 0.1250 & 1.8230(38) & 1.9470(51) & 0.9363(9) \\
1.90 & 0.1300 & 1.5753(39) & 1.7283(48) & 0.9115(12) \\
1.90 & 0.1325 & 1.4262(41) & 1.5979(46) & 0.8926(15) \\
1.90 & 0.1350 & 1.2669(39) & 1.4550(52) & 0.8707(20) \\
1.90 & 0.1375 & 1.0867(33) & 1.3126(52) & 0.8279(30) \\
1.90 & 0.1400 & 0.8504(50) & 1.1186(76) & 0.7602(35) \\
1.90 & 0.1425 & 0.4957(54) & 0.8292(96) & 0.5977(79) \\
\hline 
1.95 & 0.1200 & 1.9684(45) & 2.0643(47) & 0.9536(7) \\
1.95 & 0.1250 & 1.7239(43) & 1.8360(53) & 0.9390(10) \\
1.95 & 0.1275 & 1.6087(38) & 1.7398(51) & 0.9247(13) \\
1.95 & 0.1300 & 1.4660(37) & 1.6155(43) & 0.9075(12) \\
1.95 & 0.1325 & 1.2945(42) & 1.4530(58) & 0.8909(16) \\
1.95 & 0.1350 & 1.1266(38) & 1.3138(40) & 0.8575(21) \\
1.95 & 0.1375 & 0.9025(39) & 1.1177(46) & 0.8075(29) \\
1.95 & 0.1390 & 0.7404(33) & 0.9789(52) & 0.7563(35) \\
1.95 & 0.1400 & 0.5988(38) & 0.8510(69) & 0.7037(52) \\
1.95 & 0.1410 & 0.4465(35) & 0.7424(93) & 0.6015(82) \\
\hline 
2.00 & 0.1150 & 2.0969(35) & 2.1699(37) & 0.9664(5) \\
2.00 & 0.1200 & 1.8787(41) & 1.9662(47) & 0.9555(7) \\
2.00 & 0.1250 & 1.6202(36) & 1.7238(44) & 0.9399(9) \\
2.00 & 0.1275 & 1.4820(39) & 1.5985(48) & 0.9271(10) \\
2.00 & 0.1300 & 1.3250(33) & 1.4574(46) & 0.9092(13) \\
2.00 & 0.1325 & 1.1537(35) & 1.3047(47) & 0.8843(21) \\
2.00 & 0.1350 & 0.9550(37) & 1.1159(49) & 0.8559(25) \\
2.00 & 0.1375 & 0.7085(34) & 0.9044(45) & 0.7834(33) \\
2.00 & 0.1385 & 0.6028(47) & 0.8291(54) & 0.7271(49) \\
2.00 & 0.1390 & 0.5260(35) & 0.7439(55) & 0.7070(41) \\
2.00 & 0.1395 & 0.4462(40) & 0.7114(65) & 0.6272(62) \\
\hline 
2.10 & 0.1100 & 2.1624(22) & 2.2109(22) & 0.9781(3) \\
2.10 & 0.1150 & 1.9414(25) & 1.9989(31) & 0.9712(5) \\
2.10 & 0.1200 & 1.6988(21) & 1.7683(24) & 0.9607(5) \\
2.10 & 0.1250 & 1.4229(17) & 1.5066(19) & 0.9444(8) \\
2.10 & 0.1300 & 1.1023(23) & 1.2044(30) & 0.9152(11) \\
2.10 & 0.1325 & 0.9213(26) & 1.0429(37) & 0.8834(17) \\
2.10 & 0.1340 & 0.7797(41) & 0.9059(46) & 0.8606(25) \\
2.10 & 0.1350 & 0.7021(22) & 0.8453(30) & 0.8307(26) \\
2.10 & 0.1365 & 0.5410(27) & 0.7152(42) & 0.7564(48) \\
2.10 & 0.1375 & 0.4219(39) & 0.6158(51) & 0.6851(51) \\
\hline 
2.20 & 0.1100 & 2.0404(21) & 2.0765(22) & 0.9826(3) \\
2.20 & 0.1200 & 1.5588(21) & 1.6129(24) & 0.9665(6) \\
2.20 & 0.1225 & 1.4123(28) & 1.4695(31) & 0.9611(8) \\
2.20 & 0.1250 & 1.2666(25) & 1.3323(29) & 0.9507(9) \\
2.20 & 0.1275 & 1.1077(22) & 1.1788(26) & 0.9397(11) \\
2.20 & 0.1300 & 0.9289(26) & 1.0167(38) & 0.9137(21) \\
2.20 & 0.1325 & 0.7456(33) & 0.8442(34) & 0.8831(29) \\
2.20 & 0.1350 & 0.5014(31) & 0.6307(55) & 0.7950(51) \\
\hline 
2.30 & 0.1100 & 1.9371(18) & 1.9655(20) & 0.9855(2) \\
2.30 & 0.1150 & 1.6957(20) & 1.7314(22) & 0.9794(4) \\
2.30 & 0.1175 & 1.5721(20) & 1.6102(23) & 0.9763(4) \\
2.30 & 0.1200 & 1.4292(20) & 1.4692(23) & 0.9728(5) \\
2.30 & 0.1225 & 1.2959(20) & 1.3397(19) & 0.9673(5) \\
2.30 & 0.1250 & 1.1539(29) & 1.2074(29) & 0.9556(11) \\
2.30 & 0.1275 & 0.9890(25) & 1.0495(28) & 0.9424(10) \\
2.30 & 0.1300 & 0.8117(25) & 0.8812(30) & 0.9211(13) \\
2.30 & 0.1325 & 0.6297(29) & 0.7110(47) & 0.8857(29) \\
2.30 & 0.1340 & 0.4867(33) & 0.5932(40) & 0.8205(51) \\
2.30 & 0.1350 & 0.3807(36) & 0.5080(53) & 0.7493(69) \\
\end{tabular}
\end{center}
\caption{Results for $m_{\rm PS} a,$ $m_{\rm V} a$ and $m_{\rm PS}/m_{\rm V}$ 
on a zero temperature lattice of size $16^4$.
}
\label{tab:mass}
\end{table}

\begin{table}[hbt]
\begin{center}
  \begin{tabular}{lllllll}
 $\beta$ & $K_t(N_t=4)$ & $m_{\rm PS}/m_{\rm V}$ &  
$T_{pc}/m_{\rm V}$ & $K_t(N_t=6)$ & $m_{\rm PS}/m_{\rm V}$ &  
$T_{pc}/m_{\rm V}$ \\ \hline
1.600 & 0.1543(10)  & 0.346(153) & 0.217(11) & & & \\
1.650 & 0.1533(10)  &            &            & & & \\
1.700 & 0.1510(10)  & 0.396(170) & 0.234(17) & & & \\
1.800 & 0.1445(14)  & 0.690(92)  & 0.211(15) & & & \\
1.850 & 0.14019(18) & 0.7905(60) & 0.1917(20) & & & \\
1.900 & 0.13621(15) & 0.8525(39) & 0.1801(12) & & & \\
1.925 & 0.13417(23) &            &            & & & \\
1.950 & 0.13040(97) & 0.9051(64) & 0.1572(62) & & & \\
2.000 & 0.12371(73) & 0.9450(36) & 0.1398(29) & 0.13861(21) & 0.725(16) & 0.2086(53) \\
2.100 & 0.10921(43) & 0.9790(13) & 0.1114(09) & 0.13365(40) & 0.8635(78) & 0.1753(58) \\
2.200 & & & & 0.12539(25) & 0.9481(19) & 0.1275(15) \\
2.300 & & & & 0.11963(15) & 0.9724(12) & 0.11145(62) \\
\end{tabular}
\end{center}
\caption{Finite temperature transition/crossover point $K_t$
for $N_t=4$ and $6$. 
Results for $m_{\rm PS}(T=0)/m_{\rm V}(T=0)$ 
and $T_{pc}/m_{\rm V}(T=0)$ interpolated to the $K_t$ point 
are also listed.
}
\label{tab:Kt}
\end{table}

\begin{table}[hbt]
\begin{center}
  \begin{tabular}{llrrr}
$\beta$ & $K$ & \multicolumn{3}{c}{$-\frac{1}{N_{s}^{3} N_{t}} 
\left\langle \frac{\partial S}{\partial K} \right\rangle$} \\
     &        & $16^3 \times 4$ & $16^3 \times 6$ & $16^4$ \\ \hline
1.80 & 0.1300 & $-$5.3622(47)  &              & $-$5.4232(59) \\
1.80 & 0.1350 & $-$4.8166(79)  &              & $-$4.9341(51) \\
1.80 & 0.1375 & $-$4.3207(63)  &              & $-$4.5143(54) \\
1.80 & 0.1400 & $-$3.5759(65)  &              & $-$3.9051(93) \\
1.80 & 0.1425 & $-$2.2016(162) &              & $-$2.9788(112) \\
1.80 & 0.1440 & $-$0.2992(227) &              & $-$2.1623(102) \\
1.80 & 0.1450 &  2.1328(221) &              & $-$1.4493(102) \\
\hline 	     
1.85 & 0.1250 & $-$4.7935(43)  &              & $-$4.8429(33) \\
1.85 & 0.1300 & $-$4.3596(47)  &              & $-$4.4437(33) \\
1.85 & 0.1350 & $-$3.4384(56)  &              & $-$3.6678(50) \\
1.85 & 0.1375 & $-$2.6171(80)  &              & $-$3.0360(71) \\
1.85 & 0.1400 & $-$0.8984(161) &              & $-$2.0797(142) \\
1.85 & 0.1425 &  2.6710(146) &              & $-$0.6809(104) \\
1.85 & 0.1440 &  4.5010(140) &              &  0.7871(132) \\
\hline 	     
1.90 & 0.1250 & $-$3.9170(71)  &              & $-$3.9860(32) \\
1.90 & 0.1300 & $-$3.2151(80)  &              & $-$3.3875(53) \\
1.90 & 0.1325 & $-$2.6335(101) &              & $-$2.9136(34) \\
1.90 & 0.1350 & $-$1.5613(124) &              & $-$2.2669(69) \\
1.90 & 0.1375 &  0.3121(130) &              & $-$1.3867(70) \\
1.90 & 0.1400 &  2.6656(93)  &              & $-$0.0358(137) \\
1.90 & 0.1425 &  4.9308(89)  &              &  2.0916(163) \\
\hline 	     
1.95 & 0.1200 & $-$3.5434(50)  &              & $-$3.6066(38) \\
1.95 & 0.1250 & $-$2.9842(63)  &              & $-$3.1191(60) \\
1.95 & 0.1275 & $-$2.5015(123) &              & $-$2.7719(46) \\
1.95 & 0.1300 & $-$1.7833(65)  &              & $-$2.3076(73) \\
1.95 & 0.1325 & $-$0.6182(99)  &              & $-$1.6726(52) \\
1.95 & 0.1350 &  0.8561(76)  & $-$0.8309(61)  & $-$0.8373(61) \\
1.95 & 0.1375 &  2.4815(64)  &  0.3734(92)  &  0.3009(84) \\
1.95 & 0.1390 &              &  1.3834(87)  &  1.2217(86) \\
1.95 & 0.1400 &  4.2807(63)  &  2.3321(134) &  2.0463(106) \\
1.95 & 0.1410 &  5.0293(99)  &  3.6109(165) &  2.9673(135) \\
\hline 	     
2.00 & 0.1150 & $-$3.2222(42)  &              & $-$3.2710(27) \\
2.00 & 0.1200 & $-$2.7735(49)  &              & $-$2.8989(40) \\
2.00 & 0.1250 & $-$1.8085(56)  &              & $-$2.2919(32) \\
2.00 & 0.1275 & $-$1.0607(82)  &              & $-$1.8253(43) \\
2.00 & 0.1300 & $-$0.0836(73)  & $-$1.2623(54)  & $-$1.2600(49) \\
2.00 & 0.1325 &  0.9962(91)  & $-$0.5093(40)  & $-$0.5103(75) \\
2.00 & 0.1350 &  2.2580(69)  &  0.5282(49)  &  0.4820(82) \\
2.00 & 0.1375 &  3.7258(78)  &  2.0073(86)  &  1.8917(77) \\
2.00 & 0.1385 &              &  2.9849(14)  &  2.5623(111) \\
2.00 & 0.1390 &  4.6687(79)  &  3.4516(133) &  2.9676(112) \\
2.00 & 0.1395 &              &  4.0484(106) &  3.3763(124) \\
\hline 	     
2.10 & 0.0900 & $-$2.9462(25)  &              & $-$2.9600(17) \\
2.10 & 0.1000 & $-$2.8643(27)  &              & $-$2.8993(14) \\
2.10 & 0.1050 & $-$2.7054(31)  &              & $-$2.7710(25) \\
2.10 & 0.1100 & $-$2.3368(50)  &              & $-$2.5538(26) \\
2.10 & 0.1150 & $-$1.8240(60)  &              & $-$2.1962(21) \\
2.10 & 0.1200 & $-$1.0100(60)  & $-$1.6474(34)  & $-$1.6509(29) \\
2.10 & 0.1250 &  0.0614(57)  & $-$0.8130(62)  & $-$0.8219(32) \\
2.10 & 0.1300 &  1.6577(59)  &  0.5028(66)  &  0.4794(45) \\
2.10 & 0.1325 &              &  1.5077(73)  &  1.3864(71) \\
2.10 & 0.1340 &              &  2.3734(77)  &  2.0926(79) \\
2.10 & 0.1350 &  3.9084(51)  &  2.9291(67)  &  2.5742(52) \\
2.10 & 0.1365 &              &  3.9064(99)  &  3.4686(71) \\
2.10 & 0.1375 &  5.3108(39)  &  4.6615(65)  &  4.1983(60) \\
\hline 	     
2.20 & 0.0700 & $-$2.3029(21)  &              & $-$2.3022(16) \\
2.20 & 0.0800 & $-$2.4132(27)  &              & $-$2.4469(16) \\
2.20 & 0.0900 & $-$2.3649(31)  &              & $-$2.4697(21) \\
2.20 & 0.1000 & $-$2.0985(35)  &              & $-$2.3012(25) \\
2.20 & 0.1100 & $-$1.4253(42)  & $-$1.8043(21)  & $-$1.8065(20) \\
2.20 & 0.1200 & $-$0.0533(45)  & $-$0.6958(26)  & $-$0.6954(24) \\
2.20 & 0.1225 &              & $-$0.2433(30)  & $-$0.2560(41) \\
2.20 & 0.1250 &  1.1380(43)  &  0.3244(35)  &  0.2736(27) \\
2.20 & 0.1275 &              &  1.0372(42)  &  0.9297(37) \\
2.20 & 0.1300 &  2.7957(50)  &  1.9547(36)  &  1.7599(28) \\
2.20 & 0.1325 &  3.8577(40)  &  3.0423(41)  &  2.7646(49) \\
2.20 & 0.1350 &  5.0923(39)  &  4.4078(41)  &  4.0746(41) \\
2.20 & 0.1365 &  5.8931(36)  &  5.3480(43)  &  5.0113(52) \\
\hline 	     
2.30 & 0.1000 &              & $-$1.8180(16)  & $-$1.8210(21) \\
2.30 & 0.1100 &              & $-$1.2098(17)  & $-$1.2178(21) \\
2.30 & 0.1150 &              & $-$0.6777(21)  & $-$0.6944(21) \\
2.30 & 0.1175 &              & $-$0.3082(20)  & $-$0.3570(29) \\
2.30 & 0.1200 &              &  0.1051(23)  &  0.0439(28) \\
2.30 & 0.1225 &              &  0.6267(24)  &  0.5338(29) \\
2.30 & 0.1250 &              &  1.2434(23)  &  1.1185(25) \\
2.30 & 0.1275 &              &  2.0087(33)  &  1.8252(21) \\
2.30 & 0.1300 &              &  2.9362(34)  &  2.7110(40) \\
2.30 & 0.1325 &              &  4.0641(38)  &  3.8077(38) \\
2.30 & 0.1340 &              &  4.8553(29)  &  4.6173(29) \\
2.30 & 0.1350 &              &  5.4316(35)  &  5.1741(50) \\
2.30 & 0.1355 &              &  5.7271(42)  &  5.4944(29) \\
2.30 & 0.1360 &              &  6.0362(32)  &  5.8475(45) \\
\end{tabular}
\end{center}
\caption{Derivative $-\frac{1}{N_{s}^{3} N_{t}} 
\left\langle \frac{\partial S}{\partial K} \right\rangle$
for $16^3 \times 4, 16^3 \times 6$ and $16^4$ lattices. 
}
\label{tab:dsdk}
\end{table}

\begin{table}[hbt]
\begin{center}
  \begin{tabular}{lllll}
$\beta$ & $K$ & \multicolumn{3}{c}{$-\frac{1}{N_{s}^{3} N_{t}} 
\left\langle \frac{\partial S}{\partial \beta} \right\rangle $} \\
     &        & $16^3 \times 4$ & $16^3 \times 6$ & $16^4$ \\ \hline
1.80 & 0.1300 & 10.0029(19) &             &  9.9976(24) \\
1.80 & 0.1350 & 10.1115(21) &             & 10.1025(16) \\
1.80 & 0.1375 & 10.1788(14) &             & 10.1691(14) \\
1.80 & 0.1400 & 10.2635(11) &             & 10.2447(19) \\
1.80 & 0.1425 & 10.3846(18) &             & 10.3365(19) \\
1.80 & 0.1440 & 10.5236(20) &             & 10.4063(15) \\
1.80 & 0.1450 & 10.6747(17) &             & 10.4585(11) \\
\hline 	     
1.85 & 0.1250 & 10.2304(16) &             & 10.2282(10) \\
1.85 & 0.1300 & 10.3234(15) &             & 10.3189(12) \\
1.85 & 0.1350 & 10.4519(12) &             & 10.4365(10) \\
1.85 & 0.1375 & 10.5354(12) &             & 10.5071(16) \\
1.85 & 0.1400 & 10.6794(16) &             & 10.5974(23) \\
1.85 & 0.1425 & 10.9123(13) &             & 10.7024(13) \\
1.85 & 0.1440 & 10.9962(13) &             & 10.7981(15) \\
\hline 	     
1.90 & 0.1250 & 10.5626(27) &             & 10.5575(13) \\
1.90 & 0.1300 & 10.6646(21) &             & 10.6503(16) \\
1.90 & 0.1325 & 10.7297(23) &             & 10.7098(8) \\
1.90 & 0.1350 & 10.8326(16) &             & 10.7761(14) \\
1.90 & 0.1375 & 10.9763(14) &             & 10.8509(16) \\
1.90 & 0.1400 & 11.1143(11) &             & 10.9470(19) \\
1.90 & 0.1425 & 11.2077(11) &             & 11.0643(14) \\
\hline 	     
1.95 & 0.1200 & 10.8016(20) &             & 10.7967(16) \\
1.95 & 0.1250 & 10.8873(17) &             & 10.8763(24) \\
1.95 & 0.1275 & 10.9443(25) &             & 10.9215(16) \\
1.95 & 0.1300 & 11.0181(11) &             & 10.9724(19) \\
1.95 & 0.1325 & 11.1191(14) &             & 11.0320(13) \\
1.95 & 0.1350 & 11.2183(10) & 11.0978(12) & 11.0983(14) \\
1.95 & 0.1375 & 11.3012(9)  & 11.1747(12) & 11.1681(15) \\
1.95 & 0.1390 &             & 11.2302(11) & 11.2194(14) \\
1.95 & 0.1400 & 11.3700(9)  & 11.2747(12) & 11.2609(14) \\
1.95 & 0.1410 & 11.3929(15) & 11.3272(12) & 11.2989(12) \\
\hline 	     
2.00 & 0.1150 & 11.0363(18) &             & 11.0357(13) \\
2.00 & 0.1200 & 11.1091(12) &             & 11.0977(14) \\
2.00 & 0.1250 & 11.2205(11) &             & 11.1699(10) \\
2.00 & 0.1275 & 11.2887(14) &             & 11.2175(12) \\
2.00 & 0.1300 & 11.3605(14) & 11.2629(12) & 11.2631(14) \\
2.00 & 0.1325 & 11.4221(15) & 11.3149(8)  & 11.3168(13) \\
2.00 & 0.1350 & 11.4778(12) & 11.3783(8)  & 11.3736(12) \\
2.00 & 0.1375 & 11.5332(13) & 11.4481(9)  & 11.4451(12) \\
2.00 & 0.1385 &             & 11.4902(10) & 11.4710(13) \\
2.00 & 0.1390 & 11.5571(13) & 11.5063(10) & 11.4862(10) \\
2.00 & 0.1395 &             & 11.5281(9)  & 11.5002(12) \\
\hline 	     
2.10 & 0.0900 & 11.3777(16) &             & 11.3762(10) \\
2.10 & 0.1000 & 11.4407(14) &             & 11.4358(10) \\
2.10 & 0.1050 & 11.4793(11) &             & 11.4717(11) \\
2.10 & 0.1100 & 11.5402(12) &             & 11.5100(12) \\
2.10 & 0.1150 & 11.5992(14) &             & 11.5558(9) \\
2.10 & 0.1200 & 11.6708(15) & 11.6093(13) & 11.6077(10) \\
2.10 & 0.1250 & 11.7271(15) & 11.6670(15) & 11.6654(10) \\
2.10 & 0.1300 & 11.7937(11) & 11.7355(11) & 11.7338(9) \\
2.10 & 0.1325 &             & 11.7793(10) & 11.7709(13) \\
2.10 & 0.1340 &             & 11.8120(8)  & 11.7991(13) \\
2.10 & 0.1350 & 11.8604(11) & 11.8278(8)  & 11.8128(10) \\
2.10 & 0.1365 &             & 11.8562(12) & 11.8402(9) \\
2.10 & 0.1375 & 11.8958(8)  & 11.8759(9)  & 11.8603(8) \\
\hline 	     
2.20 & 0.0700 & 11.7472(13) &             & 11.7468(9) \\
2.20 & 0.0800 & 11.7854(18) &             & 11.7782(10) \\
2.20 & 0.0900 & 11.8419(15) &             & 11.8180(10) \\
2.20 & 0.1000 & 11.8947(14) &             & 11.8647(13) \\
2.20 & 0.1100 & 11.9626(13) & 11.9254(8)  & 11.9229(10) \\
2.20 & 0.1200 & 12.0347(12) & 11.9996(7)  & 11.9983(8) \\
2.20 & 0.1225 &             & 12.0223(7)  & 12.0216(8) \\
2.20 & 0.1250 & 12.0809(11) & 12.0477(7)  & 12.0447(8) \\
2.20 & 0.1275 &             & 12.0752(7)  & 12.0687(7) \\
2.20 & 0.1300 & 12.1265(11) & 12.1065(8)  & 12.0982(9) \\
2.20 & 0.1325 & 12.1536(10) & 12.1351(10) & 12.1249(11) \\
2.20 & 0.1350 & 12.1805(9)  & 12.1631(10) & 12.1543(7) \\
2.20 & 0.1365 & 12.1936(9)  & 12.1834(10) & 12.1724(7) \\
\hline 	     
2.30 & 0.1000 &             & 12.2137(7)  & 12.2135(9) \\
2.30 & 0.1100 &             & 12.2598(7)  & 12.2593(9) \\
2.30 & 0.1150 &             & 12.2893(7)  & 12.2887(9) \\
2.30 & 0.1175 &             & 12.3066(6)  & 12.3021(9) \\
2.30 & 0.1200 &             & 12.3234(6)  & 12.3187(9) \\
2.30 & 0.1225 &             & 12.3420(6)  & 12.3348(8) \\
2.30 & 0.1250 &             & 12.3604(7)  & 12.3550(11) \\
2.30 & 0.1275 &             & 12.3812(8)  & 12.3722(7) \\
2.30 & 0.1300 &             & 12.4034(9)  & 12.3947(9) \\
2.30 & 0.1325 &             & 12.4241(8)  & 12.4180(10) \\
2.30 & 0.1340 &             & 12.4368(8)  & 12.4337(8) \\
2.30 & 0.1350 &             & 12.4464(8)  & 12.4401(8) \\
2.30 & 0.1355 &             & 12.4497(9)  & 12.4454(8) \\
2.30 & 0.1360 &             & 12.4550(7)  & 12.4509(11) \\
\end{tabular}
\end{center}
\caption{Derivative $-\frac{1}{N_{s}^{3} N_{t}} 
\left\langle \frac{\partial S}{\partial \beta} \right\rangle $
for $16^3 \times 4, 16^3 \times 6$ and $16^4$ lattices. 
}
\label{tab:dsdb}
\end{table}

%% file: figures.tex
\begin{figure}[tb]
  \begin{center}
    \leavevmode
    \epsfxsize=12cm 
    \epsfbox{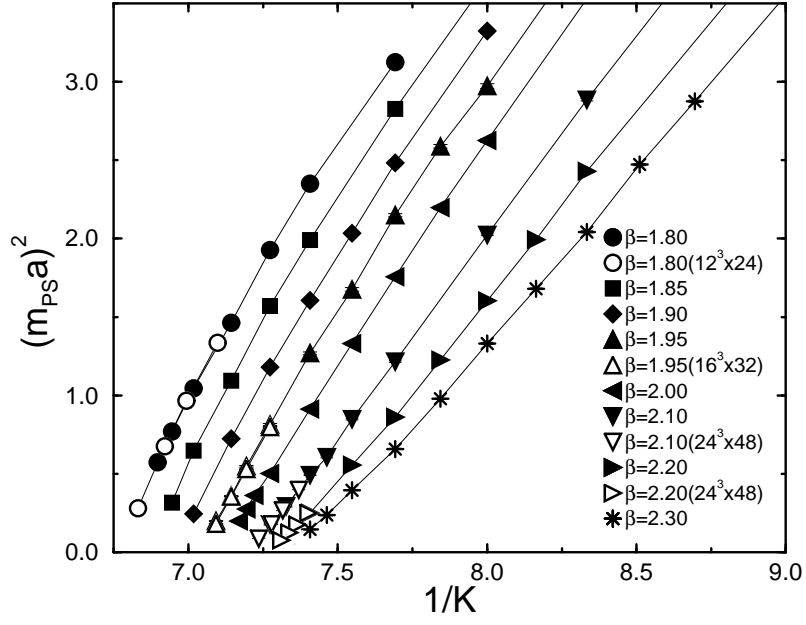}
    \caption{Pseudoscalar meson mass squared as a function of $1/K$. 
Filled symbols are obtained on a $16^4$ lattice.
Open symbols are from Ref.~\protect\cite{CPPACSmq}.}
    \label{fig:mpi2}
  \end{center}
\end{figure}

\begin{figure}[tb]
  \begin{center}
    \leavevmode
    \epsfxsize=12cm 
    \epsfbox{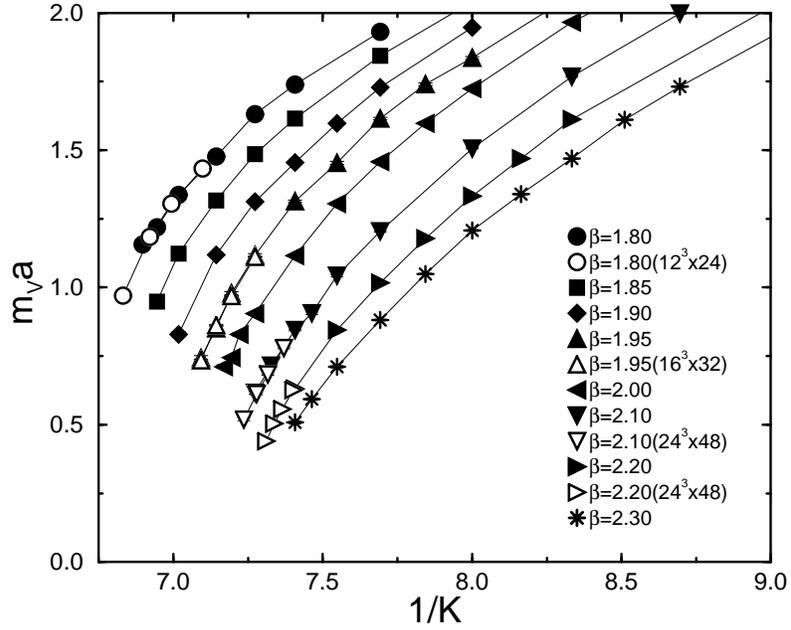}
    \caption{Same as Fig.~\protect\ref{fig:mpi2}, 
but for vector meson mass.}
    \label{fig:rho}
  \end{center}
\end{figure}

\begin{figure}[tb]
  \begin{center}
    \leavevmode
    \epsfxsize=12cm 
    \epsfbox{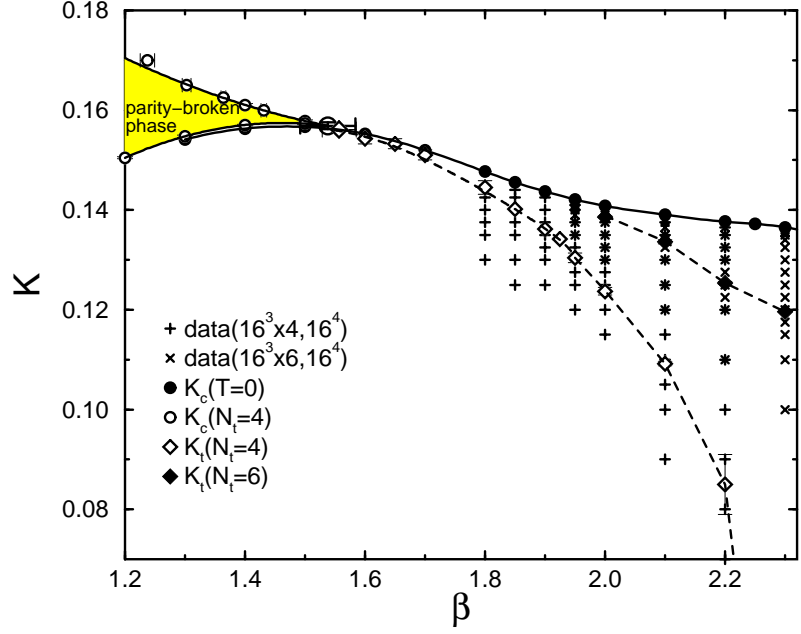}
    \caption{Phase diagram and simulation points on $16^3 \times 4$, 
	$16^3 \times 6$ and $16^4$ lattices.}
    \label{fig:phase}
  \end{center}
\end{figure}

\begin{figure}[tb]
  \begin{center}
    \leavevmode
    \epsfxsize=12cm 
    \epsfbox{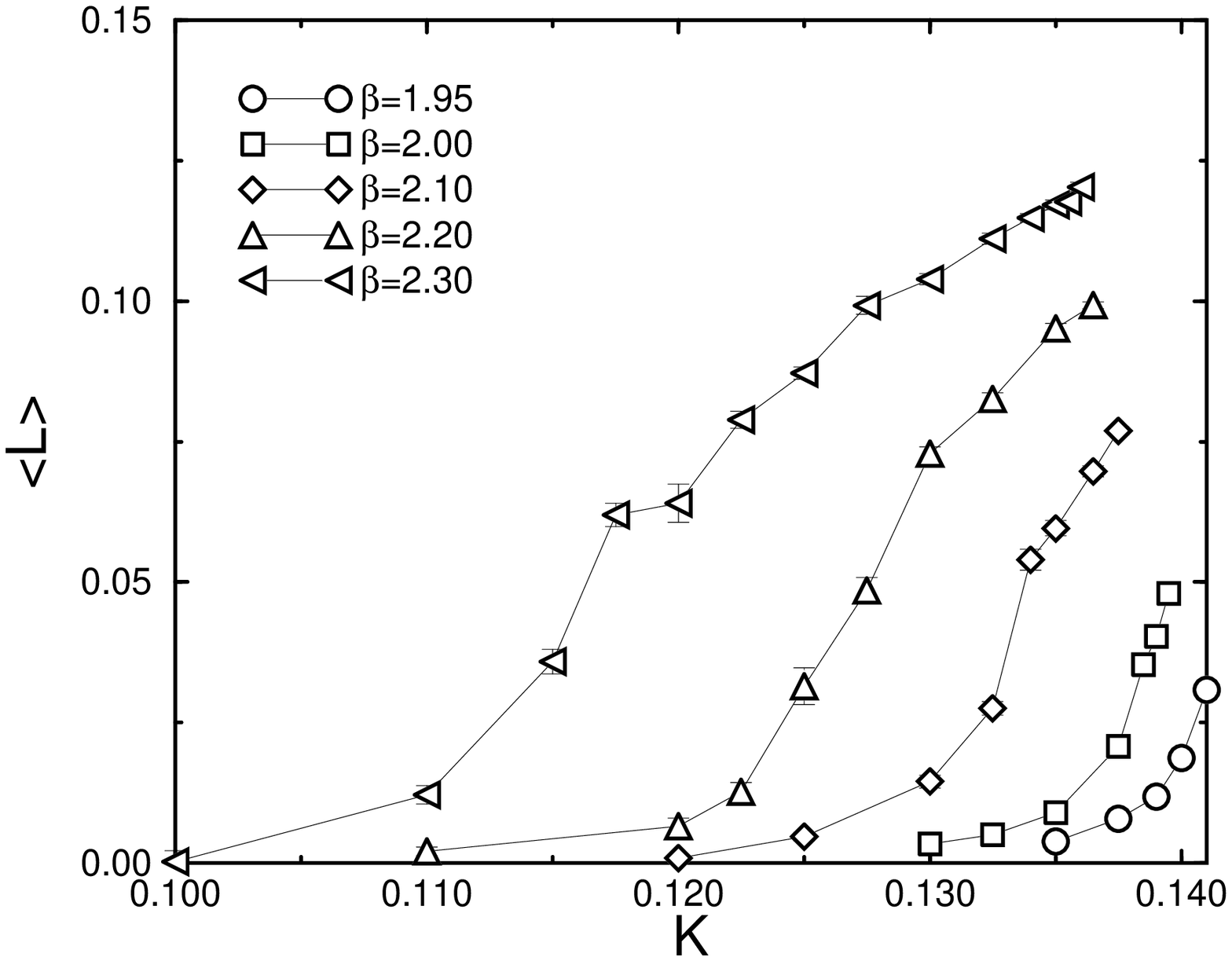}
    \caption{Polyakov loop obtained on a $16^3 \times 6$ lattice.}
    \label{fig:pline6}
  \end{center}
\end{figure}

\begin{figure}[tb]
  \begin{center}
    \leavevmode
    \epsfxsize=12cm 
    \epsfbox{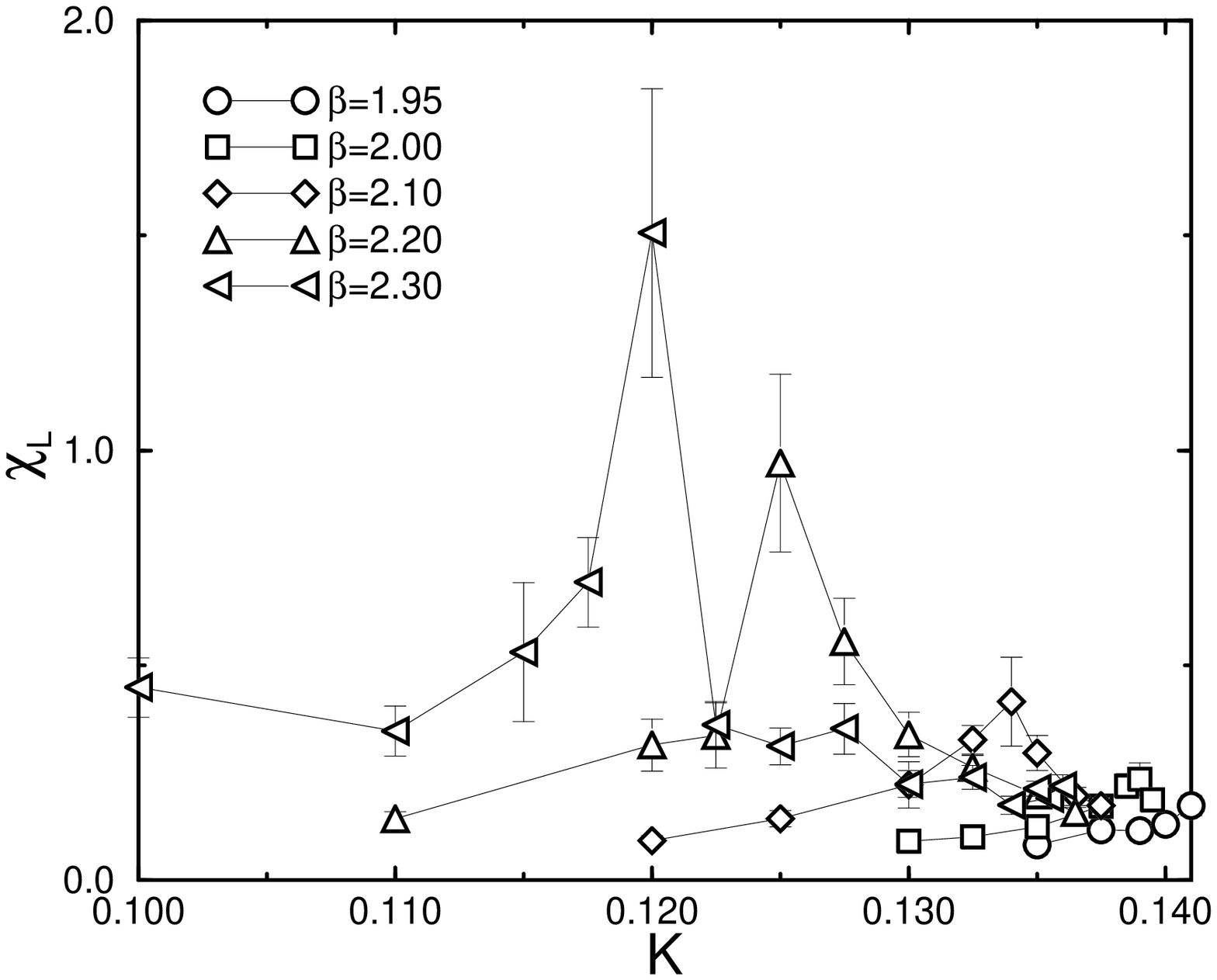}
    \caption{Polyakov loop susceptibility obtained on a 
	$16^3 \times 6$ lattice.}
    \label{fig:suspl6}
  \end{center}
\end{figure}

\begin{figure}[tb]
  \begin{center}
    \leavevmode
    \epsfxsize=12cm 
    \epsfbox{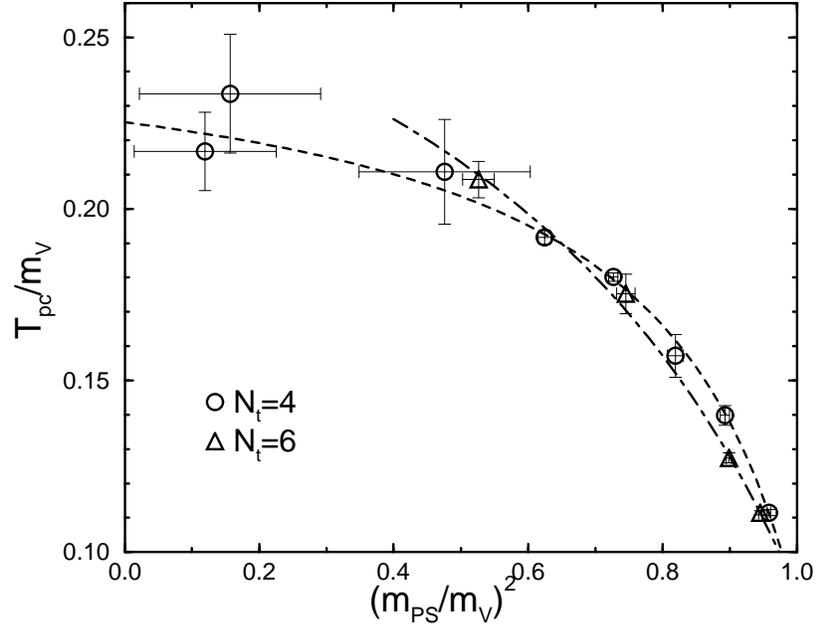}
    \caption{Pseudocritical temperature in units of vector meson mass 
as a function of $(m_{\rm PS}/m_{\rm V})^2$.
Dashed lines are interpolations, separately, for $N_t=4$ and 6 data.}
    \label{fig:tcmrho}
  \end{center}
\end{figure}

\begin{figure}[tb]
  \begin{center}
    \leavevmode
    \epsfxsize=12cm 
    \epsfbox{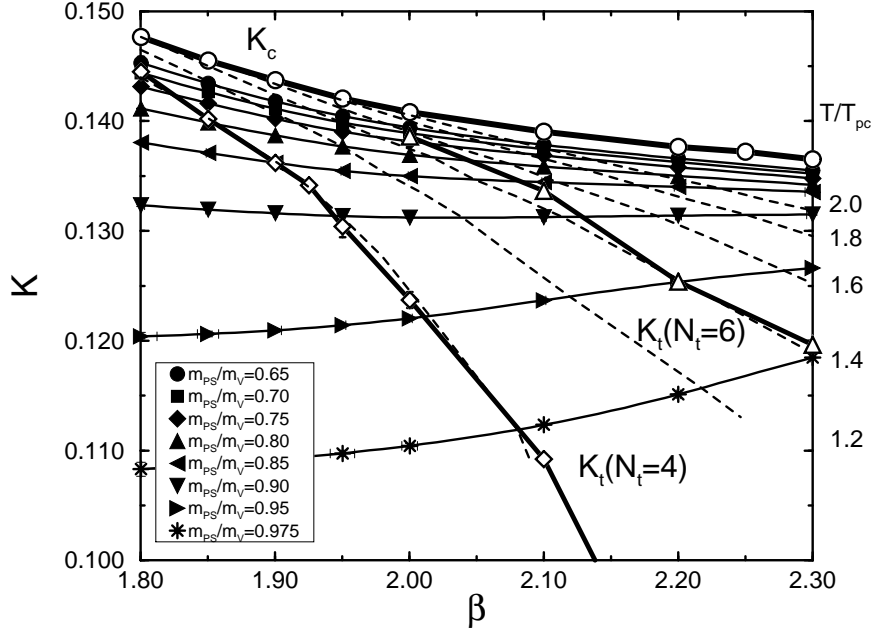}
    \caption{Lines of constant physics and of constant temperature. 
Solid lines are $m_{\rm PS}/m_{\rm V}$ constant lines, and dashed lines 
are $T/T_{pc}$ constant lines for $N_t=4$. 
The values of $T/T_{pc}$ for the dashed lines are given 
on the right edge of the figure. }
    \label{fig:Tcons}
  \end{center}
\end{figure}

\begin{figure}[tb]
  \begin{center}
    \leavevmode
    \epsfxsize=12cm 
    \epsfbox{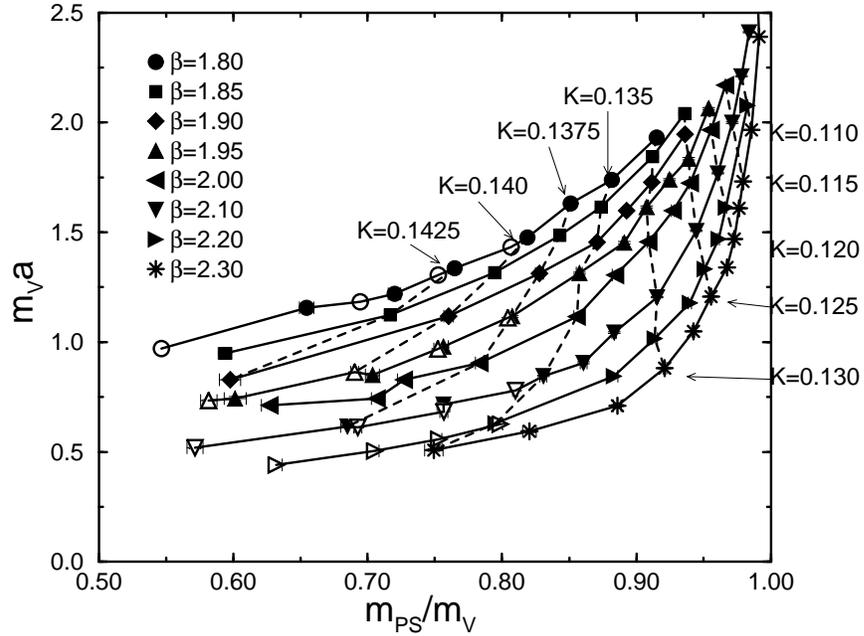} 
    \caption{$m_{\rm PS}/m_{\rm V}$ versus $m_{\rm V} a$ obtained on 
$T=0$ lattices. Solid lies are $\beta$ constant lines, and dashed lines 
are $K$ constant lines.
Meaning of symbols are the same as those in Fig.~\protect\ref{fig:mpi2}.} 
    \label{fig:pirho}
  \end{center}
\end{figure}

\begin{figure}[tb]
  \begin{center}
    \leavevmode
    \epsfxsize=12cm 
    \epsfbox{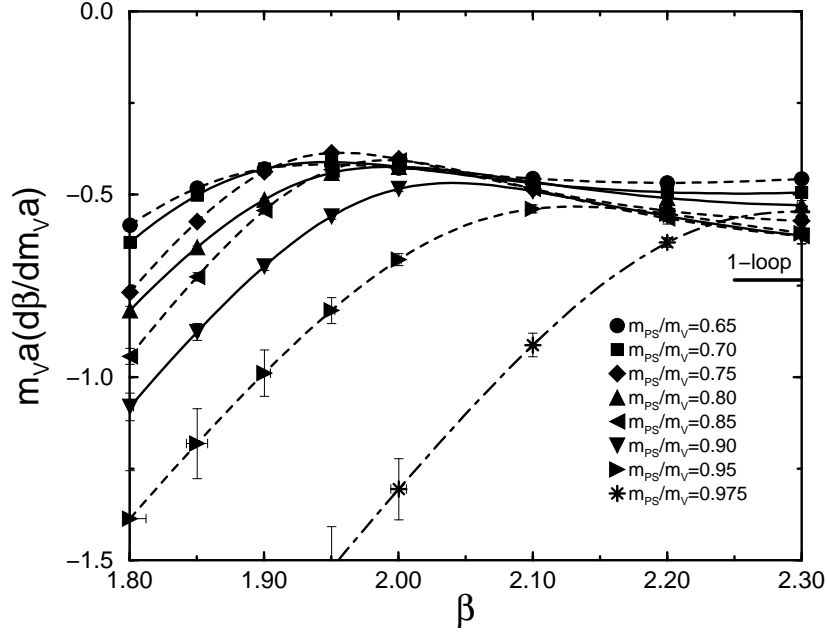} 
    \caption{$m_{\rm V}a \frac{\partial \beta}{\partial (m_{\rm V}a)}$ 
on $m_{\rm PS}/m_{\rm V}$ constant lines.
One-loop result from perturbation theory at $m_q=0$ is also shown.} 
    \label{fig:dbdrho}
  \end{center}
\end{figure}

\begin{figure}[tb]
  \begin{center}
    \leavevmode
    \epsfxsize=12cm 
    \epsfbox{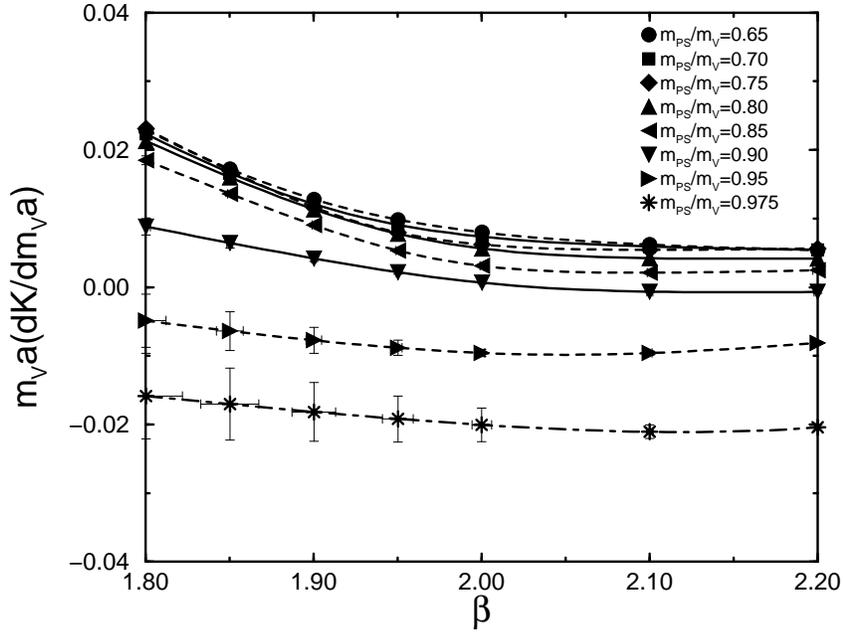} 
    \caption{$m_{\rm V}a \frac{\partial K}{\partial (m_{\rm V}a)}$ 
on $m_{\rm PS}/m_{\rm V}$ constant lines.} 
    \label{fig:dkdrho}
  \end{center}
\end{figure}

\begin{figure}[tb]
  \begin{center}
    \leavevmode
    \epsfxsize=12cm 
    \epsfbox{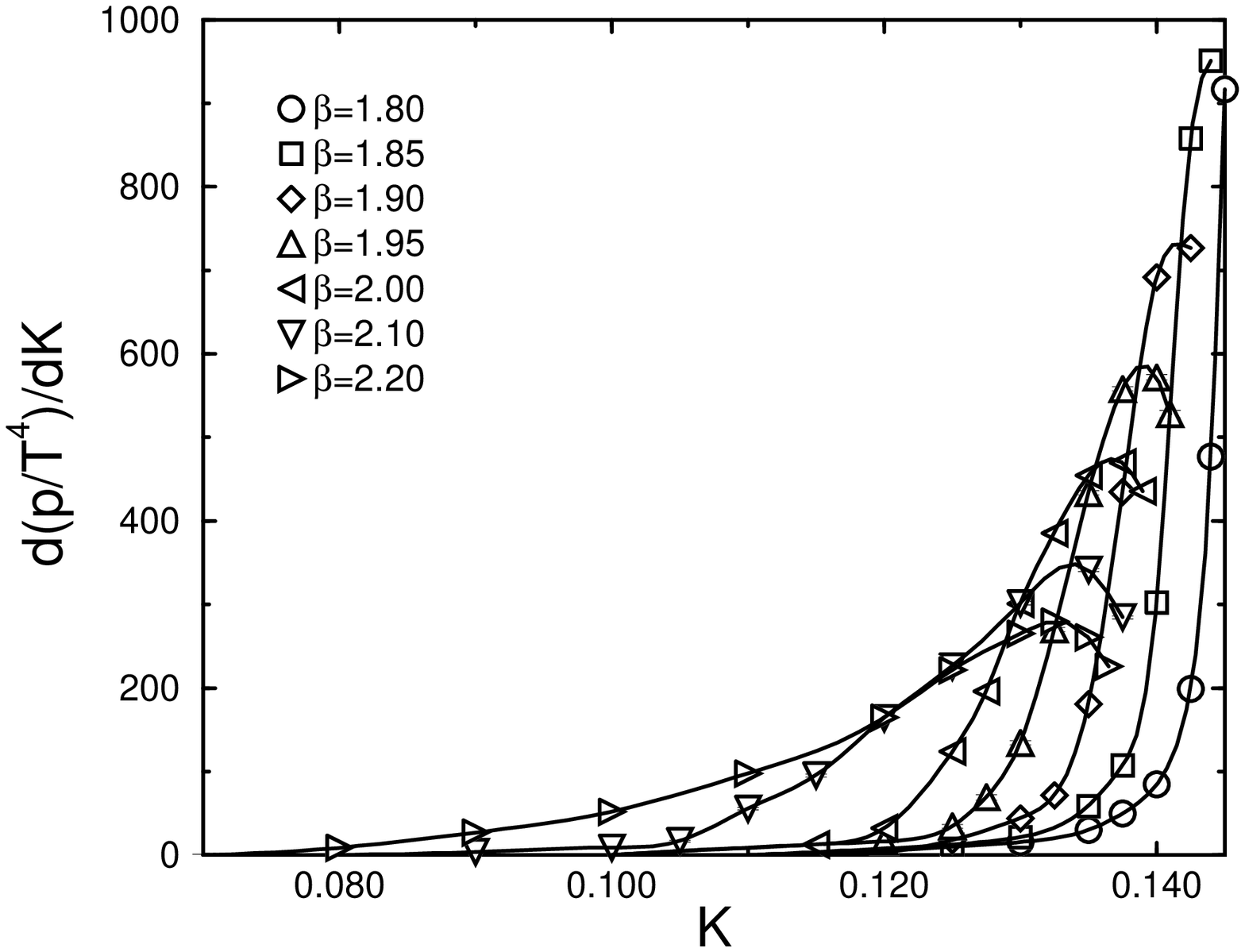} 
    \caption{Derivative of pressure with respect to $K$ 
as a function of $K$ on a $16^3\times4$ lattice.} 
    \label{fig:dpdk}
  \end{center}
\end{figure}

\begin{figure}[tb]
  \begin{center}
    \leavevmode
    \epsfxsize=12cm 
    \epsfbox{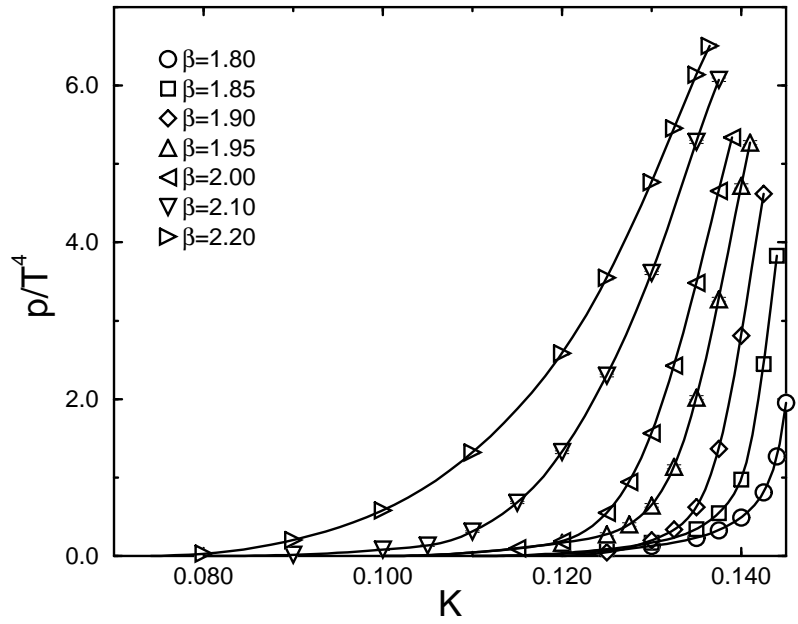} 
    \caption{Pressure as a function of $K$ on a $16^3\times4$ lattice.} 
    \label{fig:prsk}
  \end{center}
\end{figure}

\begin{figure}[tb]
  \begin{center}
    \leavevmode
    \epsfxsize=12cm 
    \epsfbox{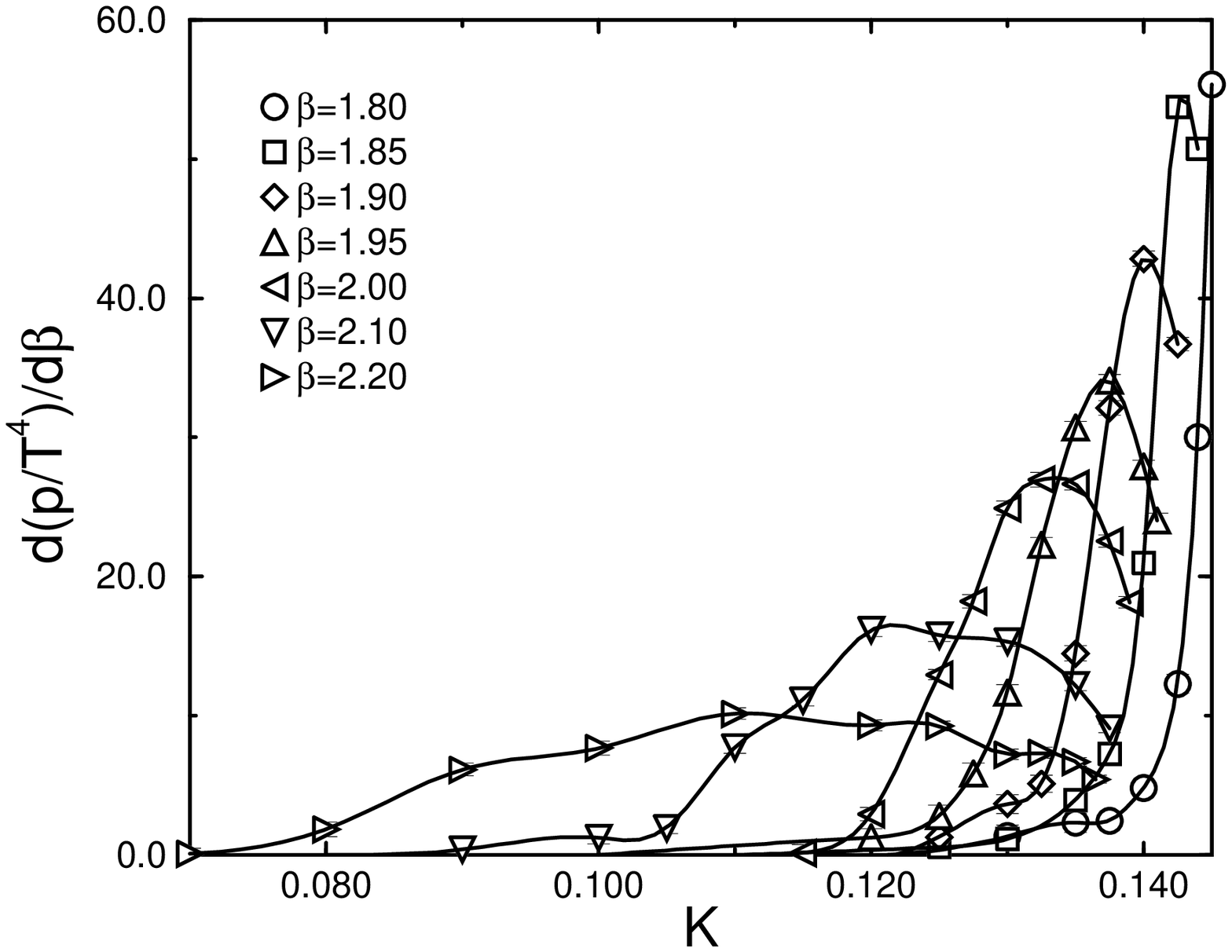} 
    \caption{Derivative of pressure with respect to $\beta$ 
    as a function of $K$ on a $16^3\times4$ lattice.} 
    \label{fig:dpdb}
  \end{center}
\end{figure}

\begin{figure}[tb]
  \begin{center}
    \leavevmode
    \epsfxsize=12cm 
    \epsfbox{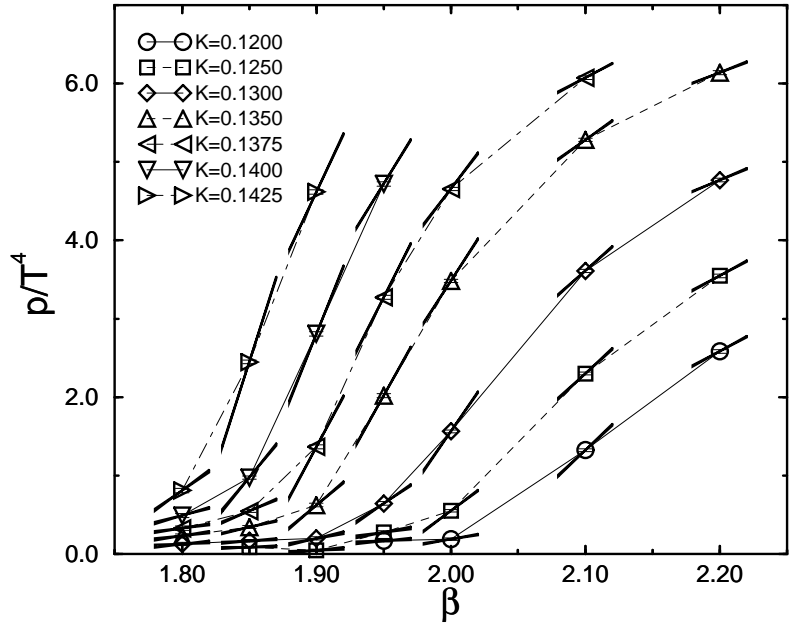} 
    \caption{Pressure as a function of $\beta$ on a $16^3\times4$ lattice.
    Short lines denote the slope of the pressure for $\beta$ direction.} 
    \label{fig:prsb}
  \end{center}
\end{figure}

\begin{figure}[tb]
  \begin{center}
    \leavevmode
    \epsfxsize=12cm 
    \epsfbox{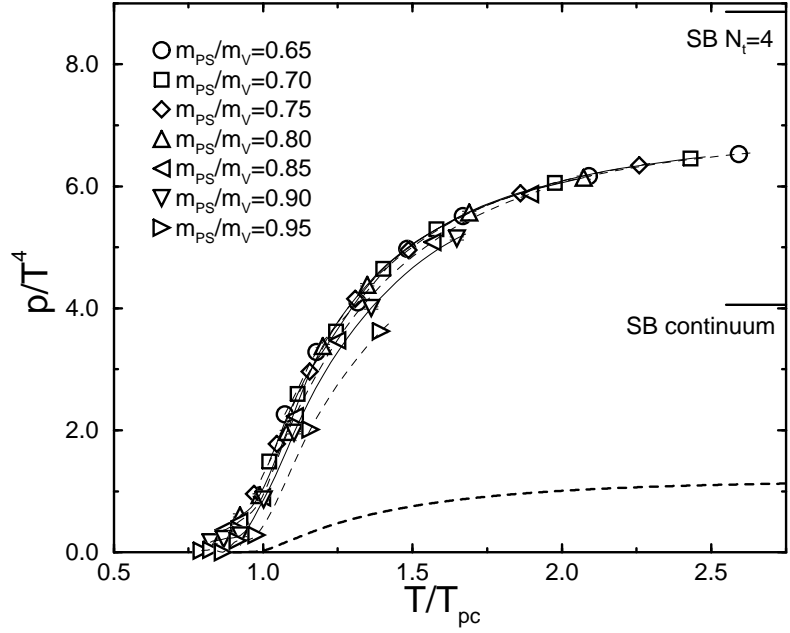} 
    \caption{Pressure on a $16^3\times4$ lattice 
    as a function of $T/T_{pc}$.
    The dashed curve shows the pressure
    for pure gauge theory with the RG-improved action 
    on a $16^3\times4$ lattice \protect\cite{okamoto}.}
    \label{fig:prsT4}
  \end{center}
\end{figure}

\begin{figure}[tb]
  \begin{center}
    \leavevmode
    \epsfxsize=12cm 
    \epsfbox{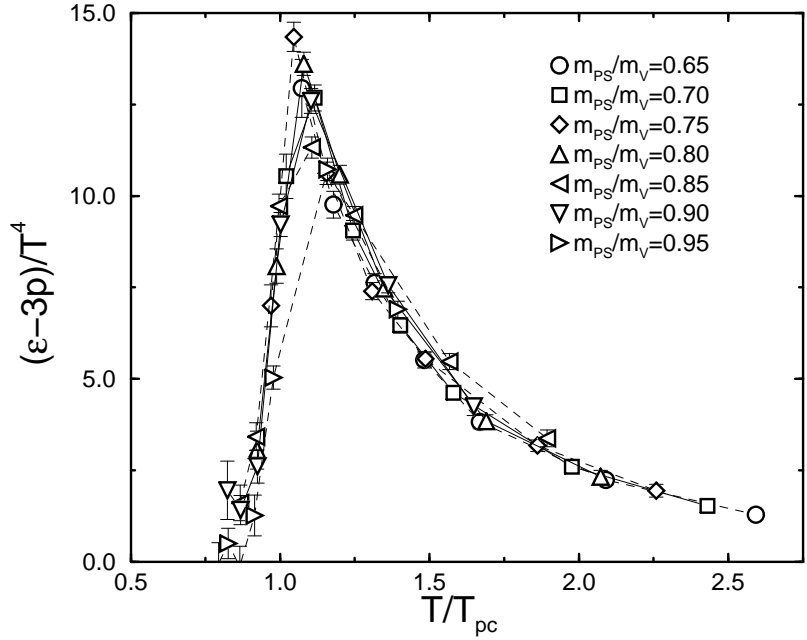} 
    \caption{$(\epsilon - 3p)/T^4$ on a $16^3\times4$ lattice 
    as a function of $T/T_{pc}$.}
    \label{fig:e3pT4}
  \end{center}
\end{figure}

\begin{figure}[tb]
  \begin{center}
    \leavevmode
    \epsfxsize=12cm 
    \epsfbox{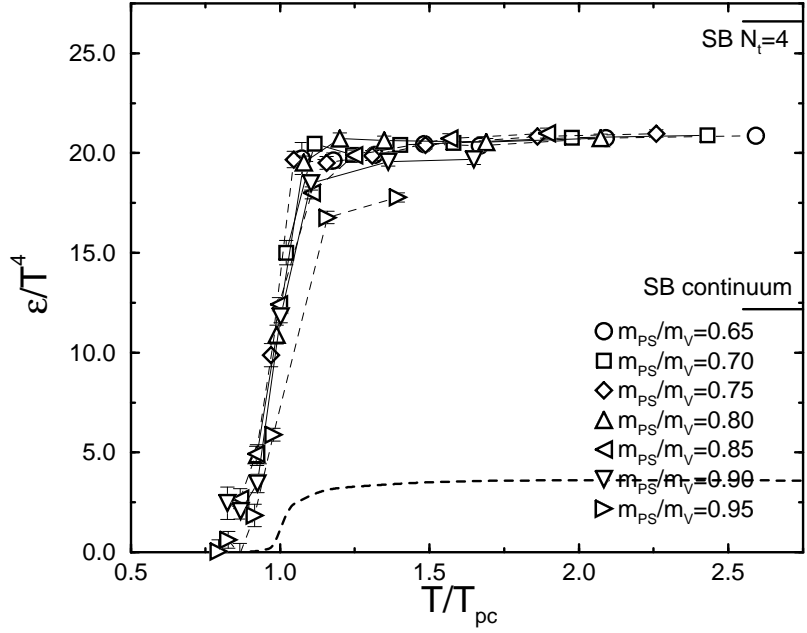} 
    \caption{Energy density on a $16^3\times4$ lattice 
    as a function of $T/T_{pc}$.
    The dashed curve shows the energy density
    for pure gauge theory with the RG-improved action 
    on a $16^3\times4$ lattice \protect\cite{okamoto}.}
    \label{fig:engT4}
  \end{center}
\end{figure}

\begin{figure}[tb]
  \begin{center}
    \leavevmode
    \epsfxsize=12cm 
    \epsfbox{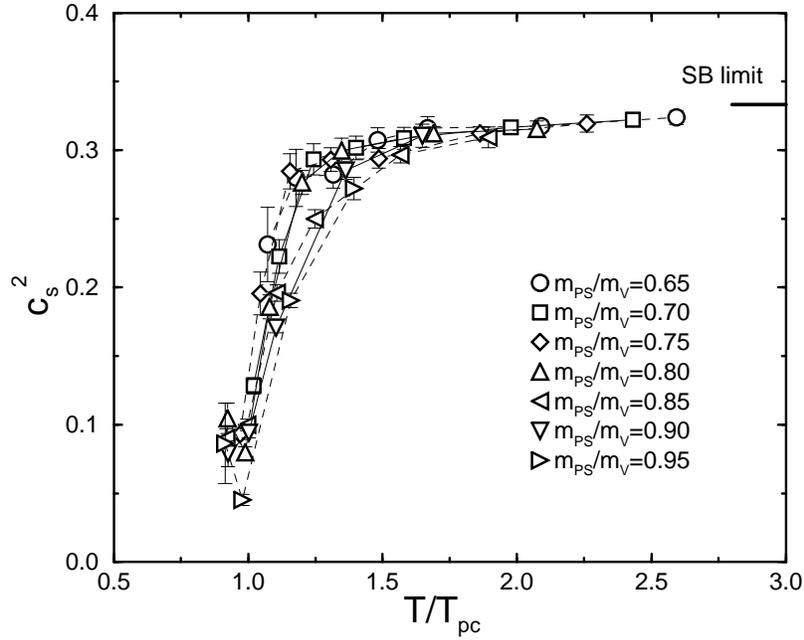} 
    \caption{Speed of sound squared on a $16^3\times4$ lattice 
    as a function of $T/T_{pc}$.}
    \label{fig:souT}
  \end{center}
\end{figure}

\begin{figure}[tb]
  \begin{center}
    \leavevmode
    \epsfxsize=12cm 
    \epsfbox{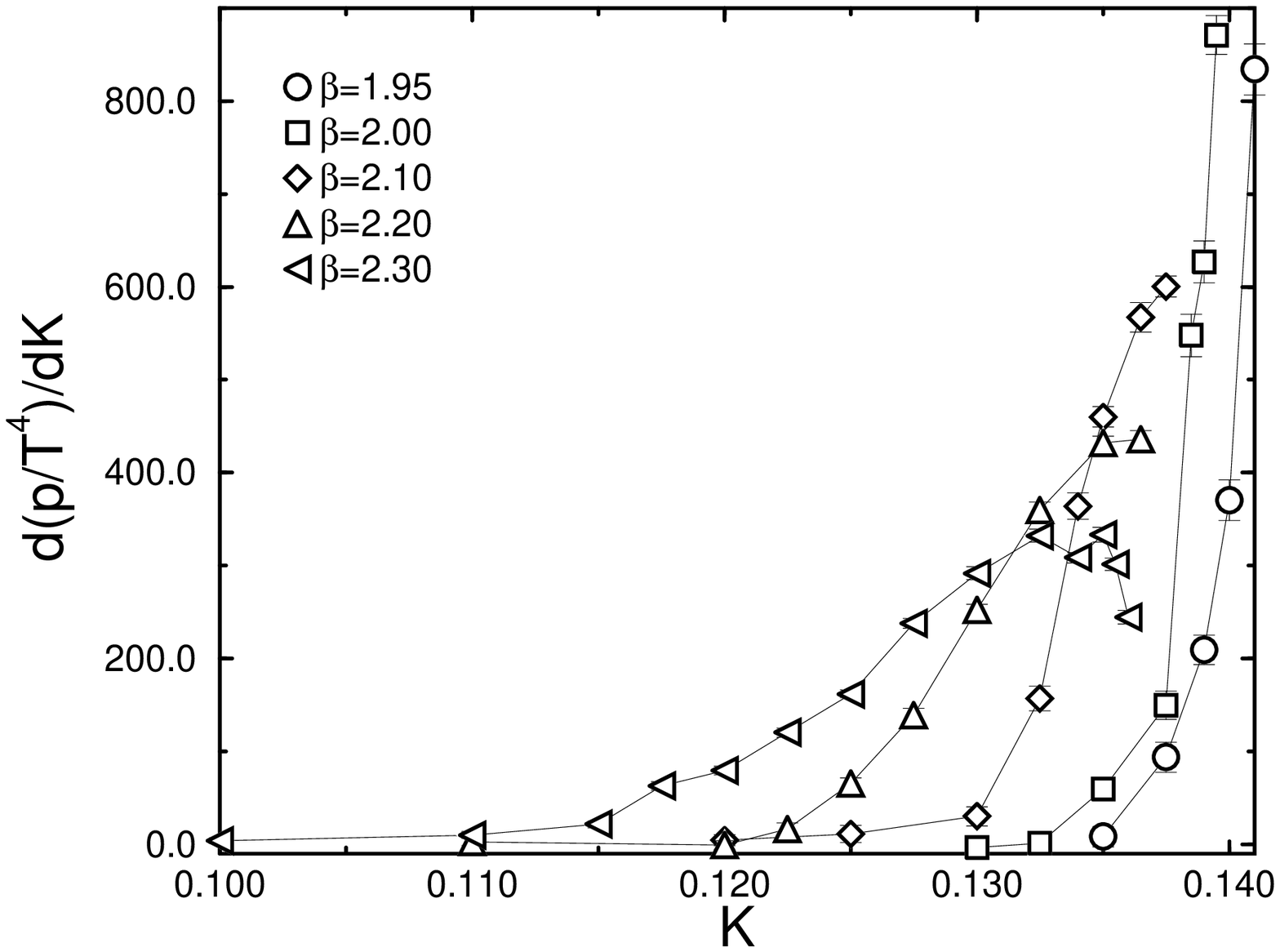} 
    \caption{Derivative of pressure with respect to $K$ 
    as a function of $K$ on a $16^3 \times 6$ lattice.} 
    \label{fig:dpdk6}
  \end{center}
\end{figure}

\begin{figure}[tb]
  \begin{center}
    \leavevmode
    \epsfxsize=12cm 
    \epsfbox{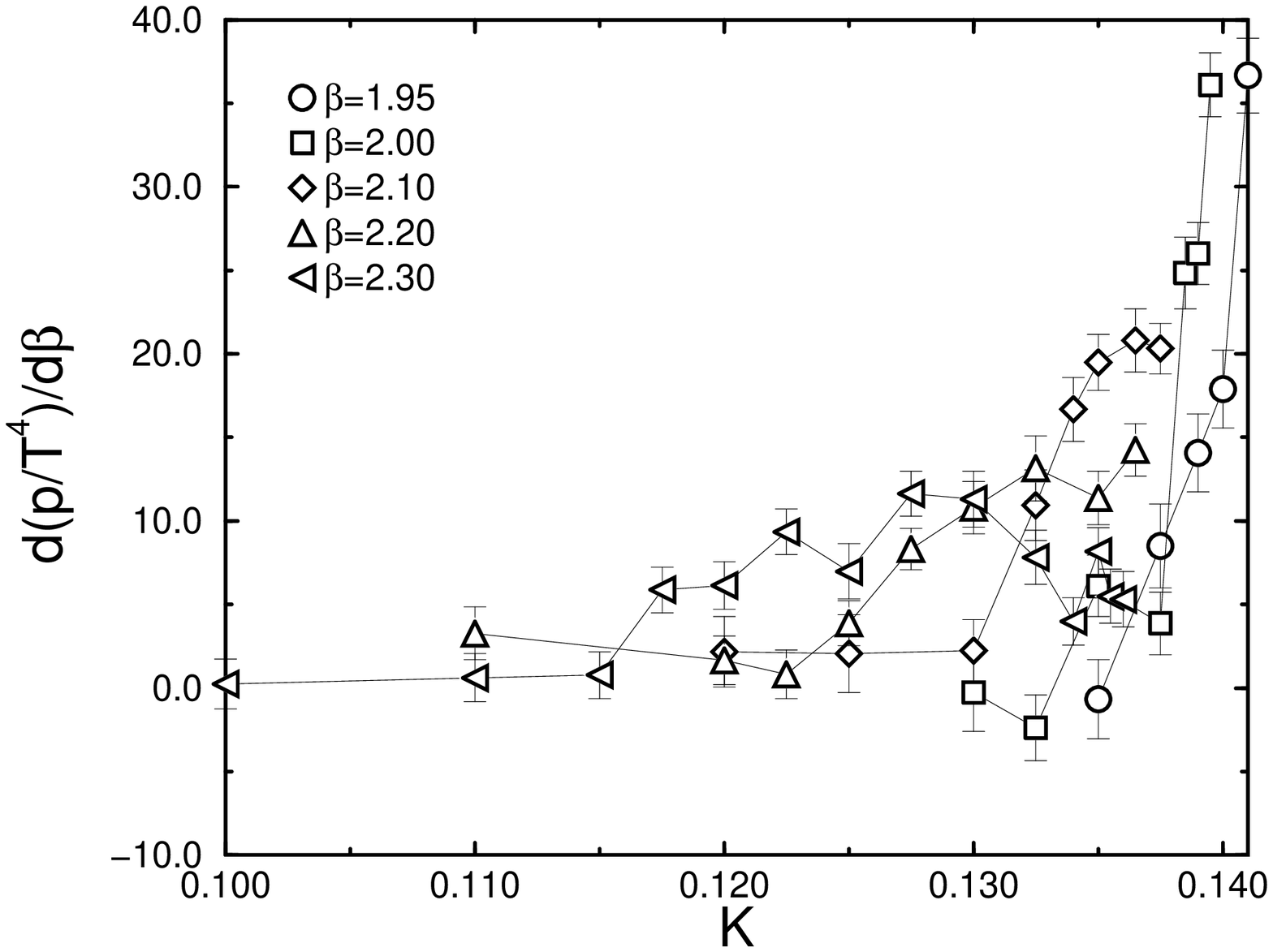} 
    \caption{Derivative of pressure with respect to $\beta$ 
    as a function of $K$ on a $16^3 \times 6$ lattice.} 
    \label{fig:dpdb6}
  \end{center}
\end{figure}

\begin{figure}[tb]
  \begin{center}
    \leavevmode
    \epsfxsize=12cm 
    \epsfbox{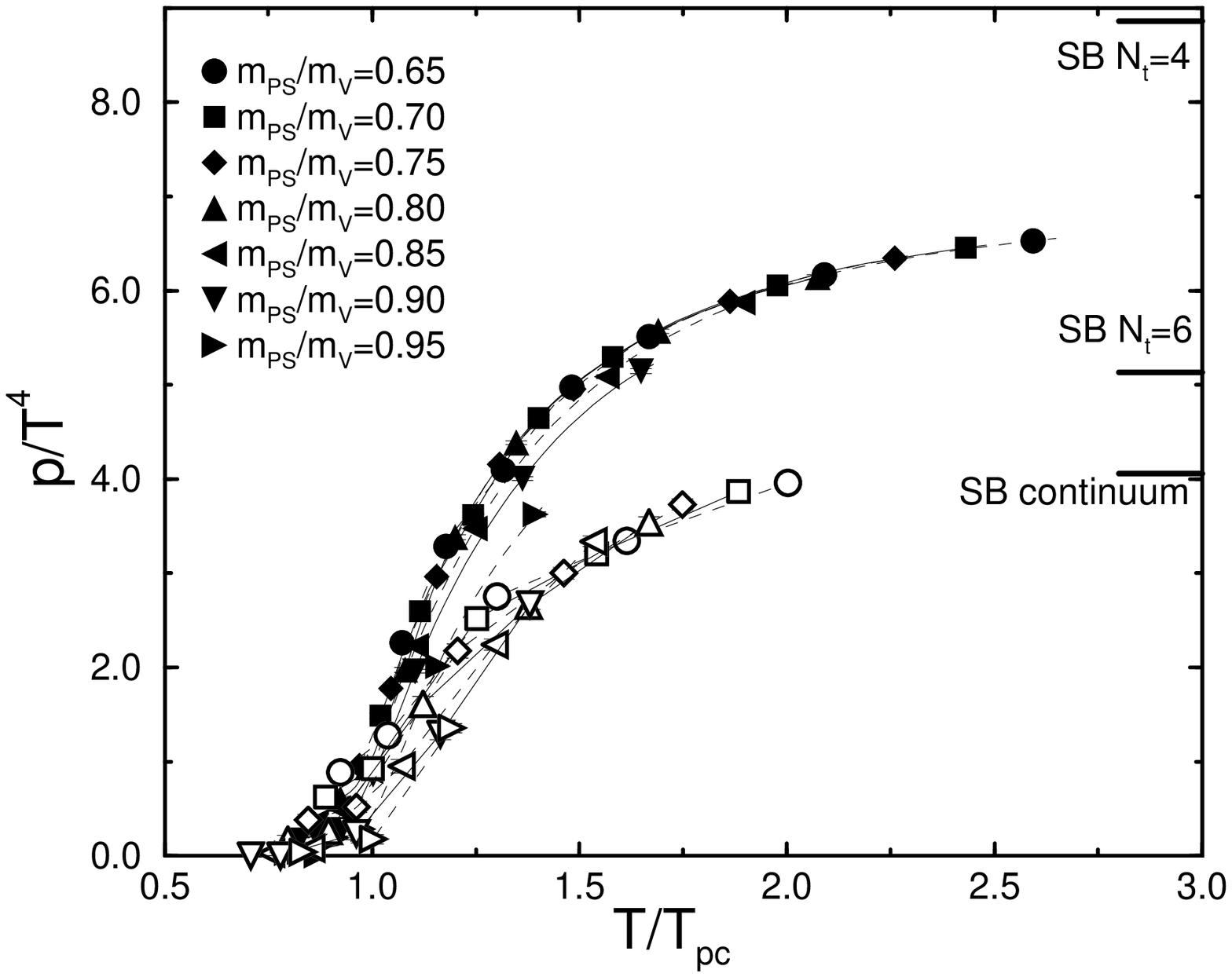} 
    \caption{Pressure on $16^3\times4$ (filled symbols) 
	and $16^3 \times 6$ (open symbols) lattices 
    as a function of $T/T_{pc}$.}
    \label{fig:prsT6}
  \end{center}
\end{figure}

\begin{figure}[tb]
  \begin{center}
    \leavevmode
    \epsfxsize=12cm 
    \epsfbox{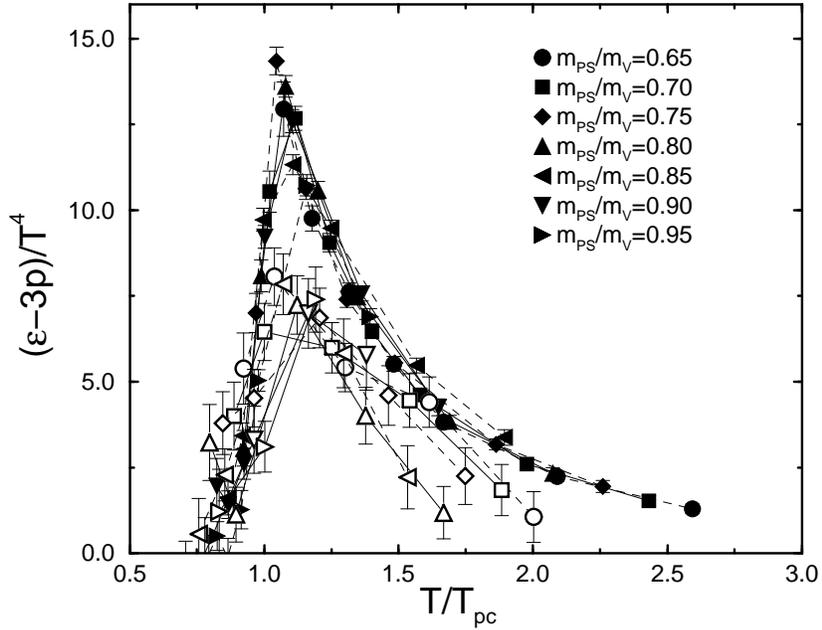} 
    \caption{$(\epsilon - 3p)/T^4$ on $16^3 \times 4$ (filled symbols) 
	and $16^3 \times 6$ (open symbols)  
    lattices as a function of $T/T_{pc}$.}
    \label{fig:e3pT6}
  \end{center}
\end{figure}

\begin{figure}[tb]
  \begin{center}
    \leavevmode
    \epsfxsize=12cm 
    \epsfbox{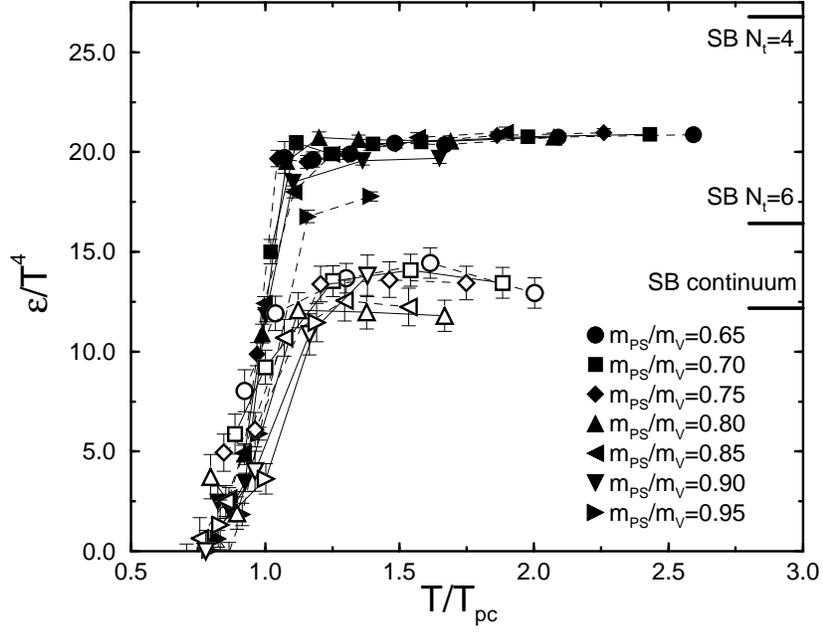} 
    \caption{Energy density on $16^3 \times 4$ (filled symbols) 
	and $16^3 \times 6$ (open symbols) lattices 
    as a function of $T/T_{pc}$.}
    \label{fig:engT6}
  \end{center}
\end{figure}

\begin{figure}[tb]
  \begin{center}
    \leavevmode
    \epsfxsize=12cm 
    \epsfbox{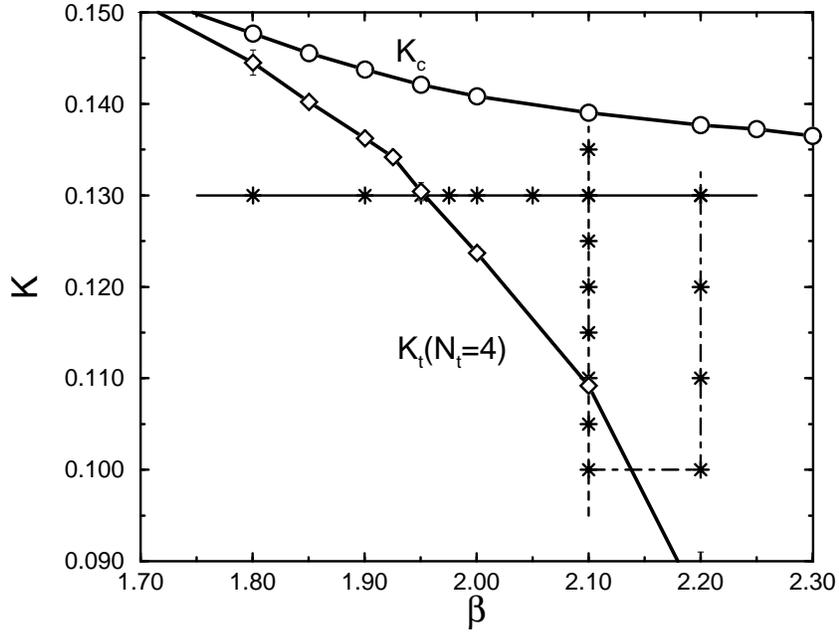} 
    \caption{Integration paths for test simulations on an $8^3 \times 4$ 
	lattice.  Stars represent the simulation points.}
    \label{fig:path84}
  \end{center}
\end{figure}

\begin{figure}[tb]
  \begin{center}
    \leavevmode
    \epsfxsize=12cm 
    \epsfbox{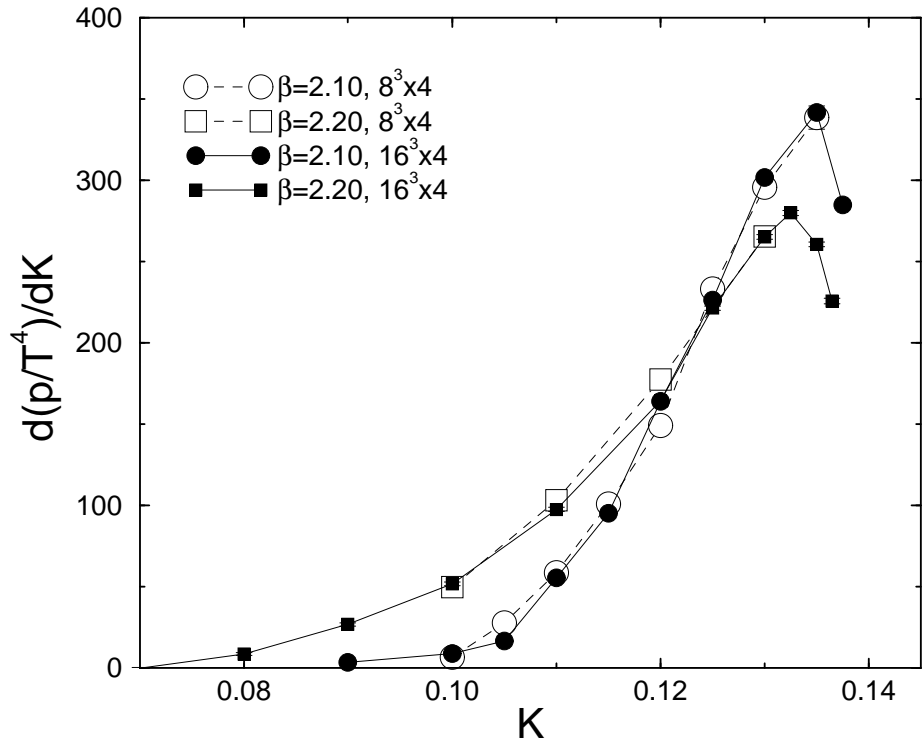}
    \caption{$\partial (p/T^4)/\partial K$ on $8^3\times4$ 
	and $16^3\times4$ lattices.}
    \label{fig:voldep_dpdk}
  \end{center}
\end{figure}

\begin{figure}[tb]
  \begin{center}
    \leavevmode
    \epsfxsize=12cm 
    \epsfbox{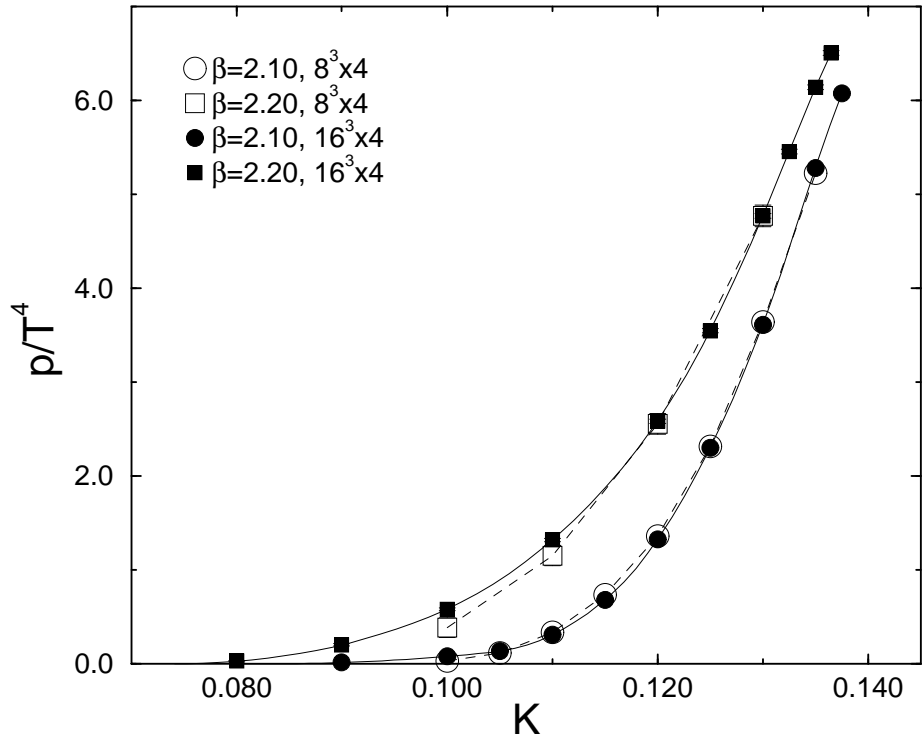}
    \caption{The pressure $p/T^4$ on $8^3\times4$ 
	and $16^3\times4$ lattices.}
    \label{fig:voldep_p}
  \end{center}
\end{figure}

\begin{figure}[tb]
  \begin{center}
    \leavevmode
    \epsfxsize=12cm 
    \epsfbox{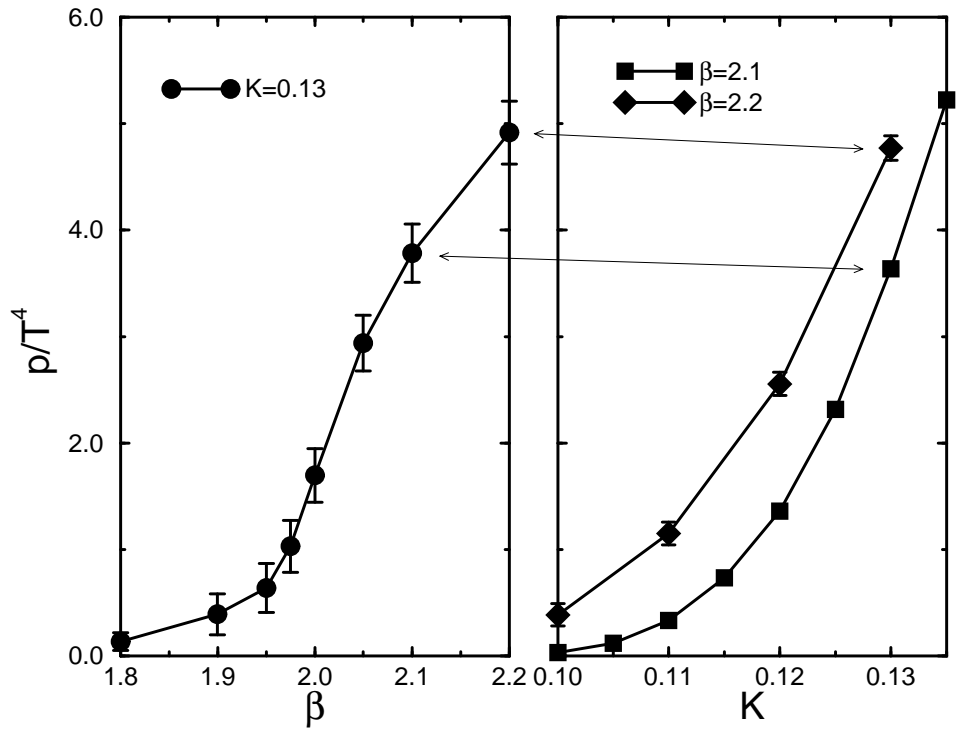} 
    \caption{Pressure computed along the integration paths at $K=0.13$ (left) 
and $\beta=2.1$ and 2.2 (right) on an $8^3 \times 4$ lattice.}
    \label{fig:prs84}
  \end{center}
\end{figure}